\documentclass[10pt]{IEEEtran}
\usepackage[pdftex]{graphicx}
\newcommand{\qed}{\fbox{}}
\newcommand{\ve}[1]{ \mbox{\boldmath$#1$} }
\newcommand{\defeq}{\stackrel{\triangle}{=}}
\newtheorem{example}{Example}
\newtheorem{definition}{Definition}

\newtheorem{lemma}{Lemma}
\newtheorem{corollary}{Corollary}

\begin{document}
\title{Interior Point Decoding for Linear Vector Channels } 
\author{Tadashi Wadayama
\thanks{T.Wadayama is with 
  Department of Computer Science, 
  Nagoya Institute of Technology, Nagoya, 466-8555, Japan.
  (e-mail:wadayama@nitech.ac.jp).
  The work was presented in part at 
  International Workshop on Statistical-Mechanical Informatics, Kyoto, Japan, Sep., 2007
  and a part of this work has been submitted to International Symposium on Information Theory, 2008.
  The initial version of this work has been  included in e-preprint server arXiv since May. 2007 
  (identificator:{\em 	arXiv:0705.3990v1}).
 } }
 \maketitle
\begin{abstract}
In this paper, a novel decoding algorithm for low-density parity-check (LDPC) 
codes based on convex optimization is presented.
The decoding algorithm, called {\em interior point decoding},  is designed for linear vector channels.
The linear vector channels include many practically important channels such as inter symbol
interference channels and partial response channels.
It is shown that the maximum likelihood decoding (MLD) rule for a linear vector channel 
can be relaxed to a convex optimization problem, which is called a {\em relaxed MLD problem}.
The proposed decoding algorithm is based on 
a numerical optimization technique so called interior point method with barrier function.
Approximate variations of the gradient descent and the Newton methods 
are used to solve the convex optimization problem. 
In a decoding process of the proposed algorithm, a search point always lies in the fundamental polytope
defined based on a low-density parity-check matrix.
Compared with a convectional joint 
message passing decoder, the proposed decoding algorithm achieves better BER 
performance with less complexity in the case of partial response channels in many cases.

\end{abstract}


\section{Introduction}

The development of decoding algorithms for binary linear codes has been 
a central research theme in coding theory.
Recent research activity on message-passing decoding algorithms has made remarkable progress, and is bringing a shift 
in the design principle of decoders from algebraic to probabilistic algorithms.
The sum-product algorithm for low-density parity-check (LDPC) codes is a particular example. The combination of LDPC codes and the sum-product algorithm 
achieves a good trade-off between decoding performance and decoding complexity.
Particularly for memoryless channels, this combination offers an almost satisfactory 
solution. 

Message passing decoding has not only been applied to memoryless channels, but also to 
channels with memory. Worthen and Stark \cite{Worthen} first presented the unified factor graph approach
for the design of a decoding algorithm for channels with memory. The unified factor graph 
includes two graphs as its sub-graphs: the factor graph for an LDPC code (the so-called Tanner graph)  and
the factor graph representing the target channel. A message passing decoding algorithm is naturally 
derived from the unified graph.  Considerable attention has been focused on this approach, with progress made on areas such as burst-error channels \cite{Garcia} and partial response channels \cite{Brian}.

The present study instead examines, the possibility of using an {\em optimization approach} for the decoding problems 
of channels with memory.  
We can view a decoding problem as an optimization problem, whereby
most conventional decoding algorithms designed for binary linear codes 
can be regarded as algorithms for solving a combinatorial optimization problem.
Let us consider the following example to make the succeeding discussion concrete. 

A codeword $\ve x$ 
in a binary code  $C \subset F_2^n$ ($F_2$ is the binary Galois field) 
is sent to an additive white Gaussian noise (AWGN) channel after binary (0,1) to bipolar (+1, -1) conversion
\begin{equation} \label{awgn}
\ve r = (\ve 1 - 2 \ve x) + \ve z,
\end{equation}
where $\ve z$ represents a Gaussian noise vector and  $\ve 1$ denotes 
the vector with all-1 elements\footnote{In this paper, the elements in $F_2$, i.e., $\{0, 1\}$,
can also be regarded as elements in ${\cal R}$. Thus, the appropriate arithmetic (e.g., mod-2 sum or 
addition of real numbers) depend on the context.}.
The maximum likelihood decoding (MLD) rule for this channel model is given by
\begin{equation}
\hat{\ve x} = \arg \min_{\ve x' \in C} || \ve r - (\ve 1 - 2 \ve x') ||^2,
\end{equation}
where $||\cdot||$ is the Euclidean norm. The rule is, indeed, a combinatorial optimization problem
because the feasible set $C$ is a discrete set with both a combinatorial and also an algebraic structure.
We often utilize  combinatorial and algebraic  properties of the code to 
solve the MLD problem.
For example, the Viterbi algorithm, which is a realization of  dynamic programming, relies heavily on 
the trellis structure of the code.  In general,  the MLD problem is 
a computationally intractable problem when the code length is large, and so
 we need   approximation algorithms that can yield sub-optimal solutions 
at a reasonable computational cost. The sum-product algorithm for LDPC codes
\cite{Gal63} can be seen as  one such approximation algorithm.

A technique called {\em relaxation} is well known 
in the field of combinatorial optimization \cite{Papa} as an approach for difficult combinatorial optimization problems. The basic idea of the relaxation is to 
relax the definition of the feasible set from a discrete set to a subset of an $n$-dimensional 
Euclidean space ${\cal R}^n$. For example, 
integer linear programming (ILP) can be regarded as a combinatorial optimization problem.
Elimination of the constraint that "a feasible point is integral" leads to a linear programming (LP) problem
that can be efficiently solved by the simplex algorithm or an interior point algorithm. 
Although the solution of the relaxed problem may be different from the
solution of the original problem, the approximate solution can be effectively used 
to find the optimal solution \cite{Papa}.

The work on LP decoding due to Feldman \cite{Feldman} is the first application of the idea of relaxation 
in coding theory. An important implication of this work on LP decoding
is that the MLD problem can be relaxed to a linear optimization problem defined on ${\cal R}^n$. 
Once a decoding problem has been relaxed to an optimization problem,  we can then  exploit an efficient numerical optimization technique to solve the relaxed problem.

Hitherto, most of the research activity on LP decoding has focused on
memoryless binary-input output-symmetric (BIOS) channels.  This may be because
the formulation of the LP problem is strongly dependent on the memoryless property and the BIOS assumption. 
 It is therefore challenging and meaningful to consider a relaxed MLD problem for channels with memory.
For example, channels with inter-symbol interference (ISI) and partial response (PR) channels are of practical importance. 
  The development of an efficient decoding algorithm for such channels is an important 
problem in coding theory.
Recently, Taghavi and Siegel \cite{Mohammad} showed that a new relaxation method for PR channels  converts
the MLD problem for PR channels to a linear programming problem.
This work indicates that the relaxation approach is effective not only for memoryless channels, but also for
channels with memory.

In this paper, a novel decoding algorithm for LDPC codes 
based on convex optimization is presented.
The decoding algorithm, called {\em interior point decoding},  is designed for linear vector channels.
The linear vector channels include many  channels of practical importance, such as ISI channels and PR channels.
It is shown that the MLD rule for a linear vector channel 
can be relaxed to a convex optimization problem, which is called a {\em relaxed MLD problem}.
Approximate variations of the gradient descent and the Newton methods 
are used to solve the convex optimization problem.

In a decoding process of the proposed algorithm, a search point always lies in the fundamental polytope \cite{Feldman} 
defined based on an LDPC matrix.
The merit function to be minimized in a decoding process 
consists of two parts: an objective function and a log-barrier function.
The objective function is the distance between
the received vector and a point in the fundamental polytope.
The log-barrier function corresponds to the constraints on the fundamental polytope, 
where this function is used so that 
the trajectory of the search points does not get close to the boundary of the fundamental polytope.
Error analysis based on the geometrical properties of a polytope is also presented.
The decision regions of a relaxed ML decoder can be characterized by normal cones
of the affine image of the fundamental polytope. 

The contents of the paper are organized as follows. In Section 2, basic notations and definitions are introduced.  Section 3 presents a geometrical view of the relaxed MLD problem for linear vector channels.
Section 4 gives an overview of the proposed algorithm. The details of the optimization methods
are explained in Section 5 (on the gradient descent method) and 6 (on the Newton method).
Section 7 includes simulation results which show the behaviors of the proposed algorithm.
Section 8 gives conclusions.

\section{Preliminaries}

In this section, we first introduce the definitions of the {\em linear vector channel} and
the MLD problem for the linear vector channels,  before 
then discussing the relaxed decoding problem.  

\subsection{Linear vector channels and MLD rule}

Suppose that $H$ is a binary $m \times n$ matrix and
$C$ is the binary linear code defined based on $H$,  
\begin{equation}
C \defeq \{\ve x \in F_2^n: H \ve x= \ve 0\}.
\end{equation}
The matrix  $H$ is a row-regular sparse matrix
whose rows have row weight $w_r \ge 3$
(In this paper, a bold-face symbol, for example $\ve x$,  denotes a column vector.).

A linear vector channel, which is the main channel considered by the present study,  is defined as follows.
\begin{definition}[Linear vector channel] 
A sender first chooses a codeword $\ve x \in C$ according to 
the message that he/she wishes to send.
The codeword is transmitted to the channel and then a 
receiver obtains a received vector $\ve r \in {\cal R}^n$:
\begin{equation} \label{linearvectorchannel}
\ve r = A \ve x + \ve b + \ve z,
\end{equation}
where  $A$ is  a non-singular $n \times n$ real matrix (called an interference matrix) and $\ve b$  is 
a real column vector of length $n$ (called an offset vector).
Note that the elements of $F_2$, $\{0,1\}$,  are also considered to be elements of ${\cal R}$ in (\ref{linearvectorchannel}). 
It is assumed that  both $A$ and $\ve b$ are known by the receiver. 
The vector $\ve z$  denotes an additive noise vector.
This channel model is called a linear vector channel.
\hfill\qed
\end{definition}

The offset vector $\ve b$ is introduced to represent   conversion of a code-representation (on $F_2$) to 
a signal-representation (on ${\cal R}$). 
For example, in the case of channel (\ref{awgn}), 
$\ve 1$ corresponds to the present $\ve b$, and is used for the binary to bipolar conversion.
Strictly speaking, the channel defined in (\ref{linearvectorchannel}) should be 
called an affine vector  channel, because the transmitted signal $\ve s$  is generated 
by an affine transformation $\ve s \defeq A \ve x + \ve b$ from a codeword $\ve x$.
However, we use the more conventional name for this channel
since the offset vector  is not essential for the following discussion,  and 
it can be seen as a part of noise or the mean of the noise. 

If the vector $\ve z$ is 
an additive white Gaussian noise vector whose $i$th element $z_i (i \in [1,n])$ has mean 0 and
variance $\sigma^2$, the channel is called a {\em Gaussian linear vector channel}.
Note that the notation $[a,b]$ denotes the set of consecutive integers from $a$ to $b$. 
 
The class of linear vector channels is wide, and includes many 
  channels of practical interest, such as the AWGN channel,
ISI  channels, and PR channels.
In order to achieve the best decoding performance with respect to the block error probability, 
we need to perform MLD
for this channel.  The following definition gives the MLD rule for
linear vector channels.
\begin{definition}[MLD rule for linear vector channel]
\label{defMLD}
Assume that the sender transmits a codeword in $C$ and 
the receiver observes $\ve r \in {\cal R}^n$ as the received word.
The MLD rule for a linear vector channel is given by
\begin{equation}\label{mldeq}
 \hat{\ve  x} = \arg \min_{\ve x \in C} d((A \ve x + \ve b),\ve r) 
\end{equation}
The function $d(\cdot, \cdot)$ is a distance function defined on ${\cal R}^n$ which matches 
the probability density function of the noise vector.
The vector $\hat{\ve  x} \in C$ is the estimation word obtained from the MLD process.
\hfill\qed
\end{definition}

\begin{example}
For the case of a Gaussian linear vector channel, 
the noise vector $\ve z$ is distributed according to an $n$-dimensional Gaussian distribution 
and its covariance matrix is diagonal (i.e., i.i.d. case).  In this case, we have the 
following MLD rule:
\begin{equation}\label{mldeqgauss}
 \hat{\ve  x} = \arg \min_{\ve x \in C} || \ve r - (A \ve x + \ve b) ||^2,
\end{equation}
where $||\cdot ||$ represents the Euclidean norm defined by
\begin{equation}
|| (x_1,x_2,\ldots,x_n) || \defeq \sqrt{x_1^2 + x_2^2 + \cdots + x_n^2}.
\end{equation}
If the noise is correlated (i.e., colored Gaussian noise), 
we can derive the MLD rule for such a channel that includes the (inverse of) covariance matrix 
of the noise\footnote{For such a channel, the appropriate distance function becomes quadratic form.}. 
\end{example}

Although the MLD rule (\ref{mldeq}) gives us an optimal estimation, its computational complexity is of an
exponential order of the code length $n$. This prevents the use of MLD in practical applications, and there is thus a requirement for  an approximation of the MLD rule in order to reduce this computational cost.

\subsection{Relaxed MLD rule}
\label{relaxedMLDrule}
We here introduce a relaxed MLD rule which is an approximation of 
the MLD rule (\ref{mldeq}). This relaxed rule is the basis of the interior point decoding 
to be presented in the latter sections. The basic idea of the relaxed rule is to 
relax the domain of $\ve x$ from $C$ to a fundamental polytope,   introduced by Feldman \cite{Feldman}, 
that is defined on the basis of the
parity check matrix $H$.
The fundamental polytope is a polytope contained  in ${\cal R}^n$ that is a relaxed 
polytope of the convex hull of $C$. Thus, 
the set of vertices of the fundamental polytope 
contains all the codewords of $C$. 
This relaxation approach thus yields the possibility of using a minimization algorithm working on ${\cal R}^n$ (e.g., a gradient descent algorithm or the 
Newton method) as a decoding algorithm.
In the following discussion,  we assume that  the distance function $d(\ve x, \ve y)$ is a convex 
function with respect to the variable $\ve x$, and that it is a differentiable function.

The definition of the fundamental polytope is given as follows\footnote{Note that although the definition of the fundamental 
polytope given here may appear at first glance to be somewhat different from that given in \cite{Feldman},   the two definitions 
are equivalent.}. 
\begin{definition}[Fundamental polytope and its interior set]
Let 
$
A_i \defeq \{j \in [1,n]: h_{ij} = 1  \},
$
for $i \in [1,m]$ where $h_{ij}$ is the
$(i,j)$-element of $H$.  The set $T_i (i \in [1,m])$ is the set of 
all the subsets of odd size in $A_i$, namely
$
T_i \defeq \{ S \subset A_i: |S| \mbox{ is odd}\}.
$
The constraints for $\ve x = (x_1,x_2,\ldots, x_n) \in {\cal R}^n$:
\begin{equation}\label{nocodeword}
\forall i \in [1,m], \forall S \in T_i,\quad 1 + \sum_{t \in S} (x_t-1) - \sum_{t \in A_i \backslash S} x_t  \le 0,
\end{equation}
and
\begin{equation}\label{box}
\forall j \in [1,n],\quad 0 \le x_j \le  1
\end{equation}
are called the parity constraints and the box constraints, respectively.
The fundamental polytope ${\cal P}$ is the polytope defined by
\begin{equation}
{\cal P} \defeq \{\ve x \in {\cal R}^n: \ve x \mbox{ satisfies both constraints (\ref{nocodeword}) and (\ref{box})}  \}.
\end{equation}
The interior set of ${\cal P}$, denoted by ${\cal P}^*$,  consists of the points satisfying
\begin{equation}\label{nocodeword2}
\forall i \in [1,m], \forall S \in T_i,\quad 1 + \sum_{t \in S} (x_t-1) - \sum_{t \in A_i \backslash S} x_t  < 0,
\end{equation}
and
\begin{equation}\label{box2}
\forall j \in [1,n],\quad 0 < x_j <  1.
\end{equation}
These constraints are also called the parity constraints and the box constraints, respectively. 
A point $\ve x \in {\cal R}^n$ is called a {\em feasible point}\footnote{In a decoding process by interior point decoding, only points in the interior set of the fundamental polytope are admissible. This is 
the reason why  we say $\ve x \in {\cal P}^*$ is a feasible point. }  iff $\ve x \in {\cal P}^*$.
\hfill\qed
\end{definition}

As described before, the set of vertices of the fundamental polytope contains 
all the codewords of $C$. Such vertices are called {\em codeword vertices}.   
Note that, in general,  the set of vertices also contains 
non-codeword vectors,   which are called {\em non-codeword vertices}.
These non-codeword vertices become the main source of 
sub-optimality in decoding performance of a decoding algorithm based on 
the fundamental polytope such as LP decoding.
The fundamental polytope ${\cal P}$ is a convex set 
defined by a set of  $m 2^{w_r-1}$ parity inequalities and $n$ box inequalities.

We are now ready to discuss a relaxation of the MLD rule for linear vector channels. 
The relaxed MLD rule based on the fundamental polytope is given by the following 
definition.

\begin{definition}[Relaxed MLD rule for linear vector channel]
Let $\ve r = A \ve x + \ve b + \ve z$ be a received vector  from a linear vector channel.
A relaxed MLD rule is defined by
\begin{equation} \label{relaxedMLD}
 \hat{\ve  x} = \arg \min_{\ve x \in {\cal P}} d((A \ve x + \ve b),\ve r).
\end{equation} 
\hfill\qed
\end{definition}
Note that, as shown in (\ref{relaxedMLD}),  the domain of $\ve x$ has changed from $C$ in the non-relaxed rule  (\ref{mldeq}) to ${\cal P}$ in the relaxed rule.
Since the fundamental polytope ${\cal P}$ is convex and the objective function  
$d((A \ve x + \ve b), \ve r)$ is  a convex function, 
the optimization problem (\ref{relaxedMLD}) can be regarded as a convex optimization problem \cite{Boyd}.
This observation motivates the use of numerical optimization techniques for convex optimization, 
such as the interior point algorithm \cite{Boyd} as a decoding algorithm.
For the case of the Gaussian linear vector channel, the relaxed MLD rule is given by
\begin{equation} \label{relaxedMLD2}
 \hat{\ve  x} = \arg \min_{\ve x \in {\cal P}} ||\ve r - (A \ve x + \ve b) ||^2.
\end{equation} 

Of course, the solution of the relaxed MLD problem  may not be the solution of the original 
MLD problem because the fundamental polytope has vertices which do not belong to $C$.
Moreover, the optimal point may not be a vertex of ${\cal P}$.
However,  this compromise on the sub-optimality of the relaxed MLD rule leads to a large   reduction in the
computational complexity of the decoding.

\section{Geometrical view of relaxed MLD problem for error analysis}

Error analysis of the relaxed MLD, which has a close relationship to the geometrical properties of the fundamental polytope,
is important to clarify the difference between the true MLD and the relaxed MLD.
In this section, geometrical properties of the relaxed MLD problem will be discussed.

\subsection{Mapped polytope}

It may be helpful for us to obtain a geometric  intuition of the relaxed MLD rule before discussing further
  details.
Let ${\cal P} \subset {\cal R}^n$ be the  fundamental polytope.
Applying the affine map $\ve x \mapsto A \ve x + \ve b$ to 
${\cal P}$, we obtain the image of ${\cal P}$:
\begin{equation}
{\cal Q} \defeq \{A \ve x + \ve b \in {\cal R}^n: \ve x \in {\cal P}\}.
\end{equation}
The set ${\cal Q}$ is also a polytope, which is called 
a {\em mapped polytope}.

\begin{figure}[htbp]
\begin{center}
\includegraphics[scale=0.5]{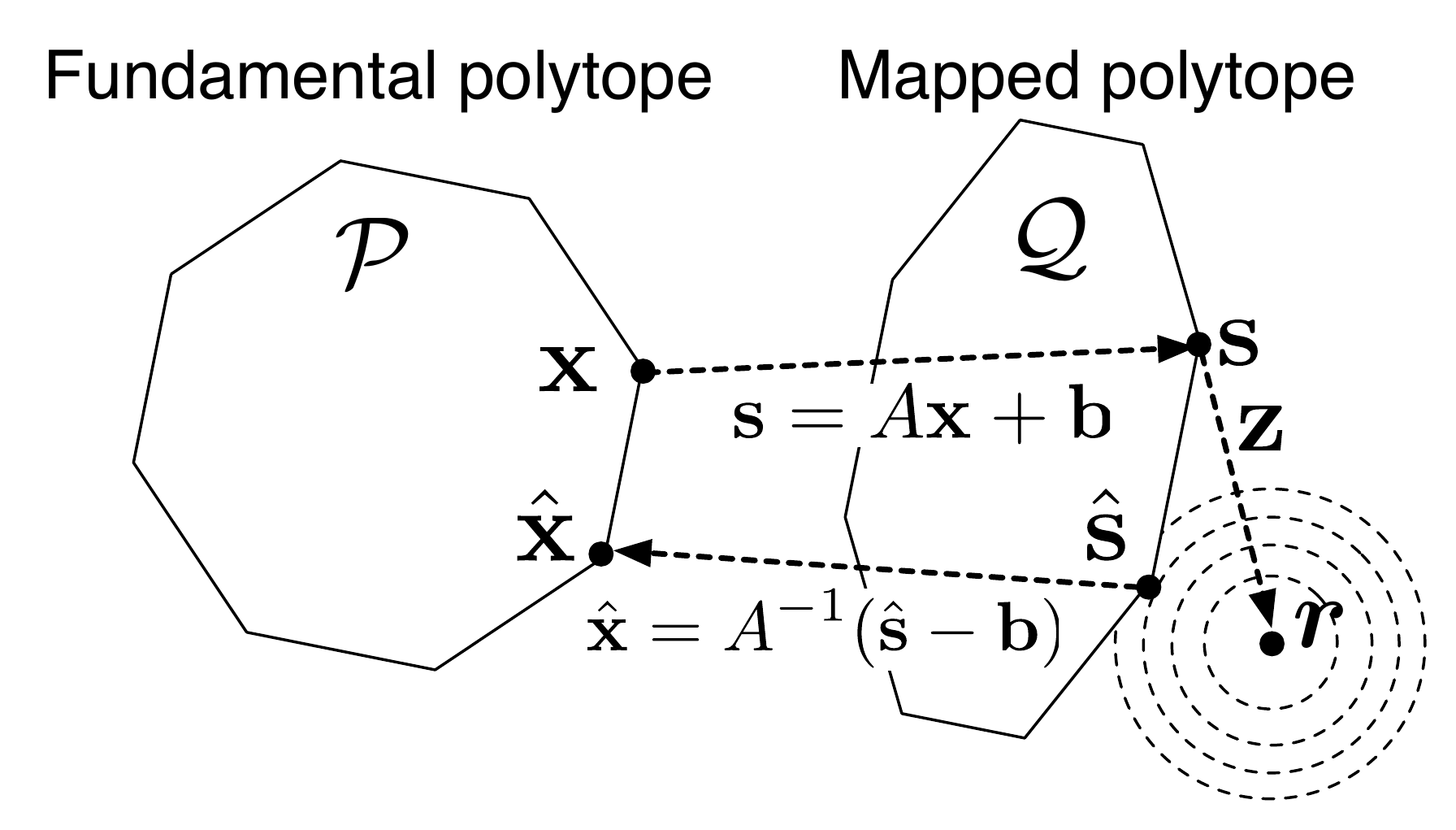} \\
  \caption{Geometrical view of the relaxed MLD process.}
  \label{fundpoly}
\end{center}
\end{figure}

By using ${\cal Q}$, we can rewrite the relaxed MLD rule in the following two-step process:
\begin{eqnarray}
\hat {\ve s} &=& \arg \min_{\ve s' \in {\cal Q}} d(\ve s', \ve r) \\
\hat{\ve x} &=& A^{-1}(\hat{\ve s} - \ve b).
\end{eqnarray}
Note that the interference matrix $A$ is assumed to be non-singular and thus the inverse $A^{-1}$ exists.
Figure \ref{fundpoly} illustrates the relation between ${\cal P},{\cal Q}, \ve r$ and $\hat{\ve x}$.
The dashed circle around $\ve r$ denotes the contours of the objective function $d(\ve s', \ve r)$. 
This figure presents  a case of mis-correction(i.e., $\ve x \ne \hat{\ve x}$).

\subsection{Normal cone}
We here review the linear vector channel model again.
The received vector $\ve r \in {\cal R}^n$ is given by
$\ve r = A \ve x + \ve b + \ve z$.
The transmitted vector $\ve x$ is a (codeword) vertex of a fundamental polytope.
The additive noise vector $\ve z \in {\cal R}^n$  is assumed to be 
generated according to the probability density function $p(\ve z)$.

In order analyze the block error performance of the relaxed ML decoder\footnote{We here assume 
the {\em optimal} relaxed ML decoder can solve the relaxed MLD problem exactly.}, 
 a necessary and sufficient condition for 
the optimal points of the relaxed MLD problem must be established.
The {\em normal cone} defined below is the essential basis of
such a necessary and sufficient condition.

\begin{definition}[Normal cone]
Let ${\cal F} \subset {\cal R}^n$ be a polytope.
Suppose that $\ve x \in {\cal F}$. The normal cone at $\ve x$ is defined by
\begin{equation}
N_{\cal F}(\ve x) \defeq \{\ve y \in {\cal R}^n: \ve y^t (\ve x' - \ve x) \le 0, \forall \ve x' \in {\cal F}\}.
\end{equation}
The vectors belonging to $N_{\cal F}(\ve x)$ are called the normal vectors of ${\cal F}$ at
$\ve x$. \hfill\qed
\end{definition}

From the definition, it is clear that the normal cone is a closed convex cone including
the origin $0^n$. 
If $\ve x$ is included in the interior set of ${\cal F}$, $N_{\cal F}(\ve x) = \{0^n \}$ holds
because there exists an $\epsilon$-ball ($\epsilon > 0$) centered at $\ve x$ which 
is totally included in ${\cal F}$. 
We next consider the case where $\ve x$ is on a facet $S$ of 
${\cal F}$ and let $\ve z$ be an orthogonal (normal) vector to $S$. 
It can be verified that $t \ve z^t (\ve x' - \ve x) \le  0$ holds for any $\ve x' \in {\cal F}$, where
$t$ is a non-negative real number. We  can show that vectors $t \ve z$ are the only vectors
that satisfy the inequality. This means that $N_{\cal F}(\ve x) = \{t \ve z: t \ge 0 \}$.
Finally \footnote{We here omit the 
case where $\ve x$ is on the ridge (or edge) of ${\cal F}$. This case is similar to the case where $\ve x$ is on 
a vertex. }, we consider the case where $\ve x$ is a vertex of the polytope ${\cal F}$. Suppose that $\ve x$ is given as the intersection of facets $S_1,S_2,\ldots S_k$. 
The vectors $\ve z_1, \ve z_2, \ldots, \ve z_k$ are the normal vectors corresponding to these facets, respectively. 
In that case, we can show that
\begin{equation}
\ve y \in \{ t_1 \ve z_1 + \cdots t_k \ve z_k: t_1,t_2,\ldots, t_k \ge 0  \}
\end{equation}
satisfies $\ve y^t (\ve x' - \ve x) \le 0$ for any $\ve x' \in {\cal F}$ and vice versa.
This leads to the following statement: 
\begin{equation}
N_{\cal F}(\ve x) =\{ t_1 \ve z_1 + \cdots t_k \ve z_k: t_1,t_2,\ldots, t_k \ge 0  \}
\end{equation}
holds if $\ve x$ is a vertex of ${\cal F}$.

\begin{figure}[htbp]
\begin{center}
\includegraphics[scale=0.5]{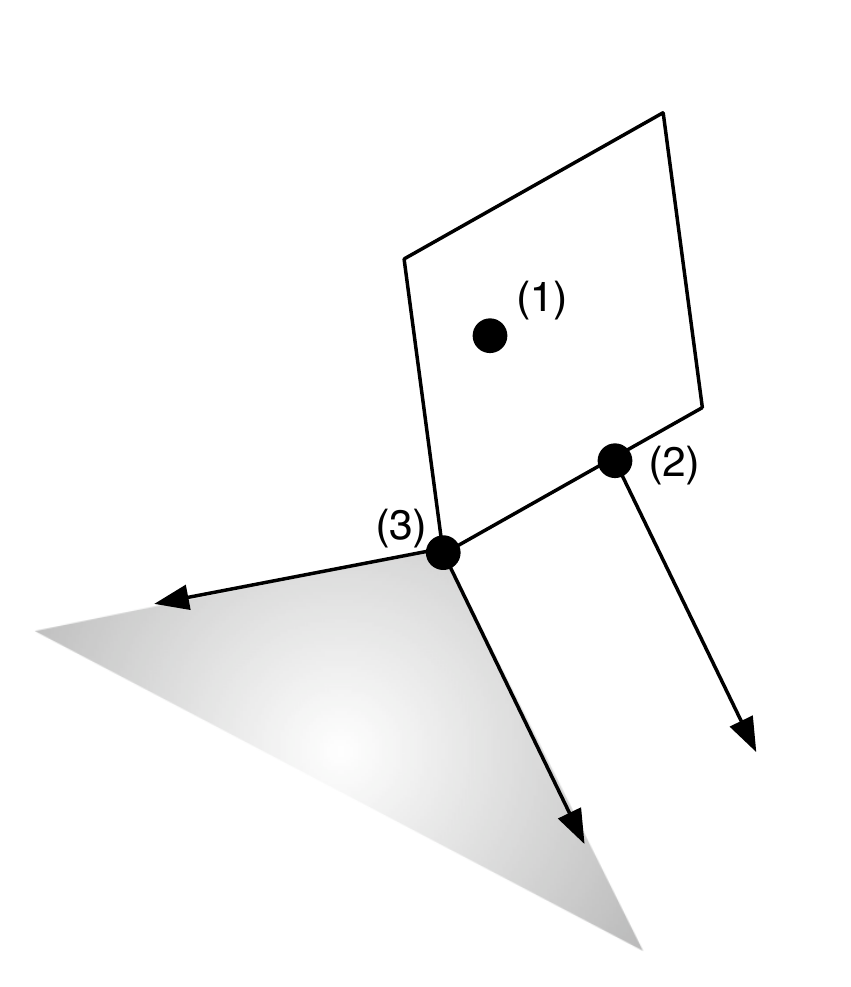} \\
  \caption{Shifted normal cones.} 
  \label{normalcone}
\end{center}
\end{figure}

The {\em shifted normal cone} $N_{\cal F}(\ve x)+ \ve x$ is defined by
\begin{equation}
N_{\cal F}(\ve x)+ \ve x \defeq \{\ve y + \ve x: \ve y \in N_{\cal F}(\ve x) \},
\end{equation}
which is a shifted cone starting from $\ve x$. Figure \ref{normalcone} illustrates 
the shifted normal cones for a two-dimensional polytope. The black circles denote
the point $\ve x$. The figure depicts the three cases discussed above ($\ve x$ is on 
(1)the interior set, (2)a facet, (3)a vertex).

\subsection{Optimality condition and decodable noise region}

The next lemma is the basis of the proof of the Karush-Kuhn-Tucker condition for convex optimization problems.
\begin{lemma}[Condition for global optimum]
\label{KKT}
Assume that ${\cal F} \subset {\cal R}^n$ is a convex set.
Let $f(\ve x)$ be a convex function to be minimized subject to the convex constraint
$\ve x \in {\cal F}$. We assume that $f(\ve x)$ is differentiable at any $\ve x \in {\cal F}$.
If and only if 
\begin{equation} \label{globaloptimumcond}
-\nabla f(\ve x^*) \in N_{\cal F}(\ve x^*)
\end{equation}
holds, $\ve x^*$ is the global minimum of the convex optimization problem. \\
(Proof) We omit the proof of this Lemma since it can be found in  standard textbooks on non-linear optimization, such as \cite{Fukushima}. 
 \hfill\qed
\end{lemma}

This lemma can be used to specify the set of decodable noise using a relaxed ML decoder.
For this purpose, it is convenient to introduce the set of decodable noise patterns.
\begin{definition}[Decodable noise region]
\label{decisionregion}
Let $\ve r \in {\cal R}^n$ be a received vector.
For any $\ve s \in {\cal Q}$,  the decodable noise region $D(\ve s)$ corresponding to $\ve s$ is
defined by
\begin{equation} 
D(\ve s) \defeq \{\ve r - \ve s \in {\cal R}^n: -\nabla d(\ve s,  \ve r) \in N_{\cal Q}(\ve s) \}.
\end{equation}
\hfill\qed
\end{definition}

The meaning of  the decodable noise region becomes clear in the following corollary:
\begin{corollary}
\label{vertexcol}
Assume that $\ve x$ is a point in the fundamental polytope ${\cal P}$.
The received vector $\ve r = A \ve x + \ve b + \ve z\in {\cal R}^n$ can be  correctly 
decoded (i.e., $\ve x = \hat{\ve x}$ holds) using the relaxed ML decoder
if  $\ve z \in D(A\ve x+\ve b)$ holds. \\
(Proof) From the assumption $\ve z = \ve r - A \ve x - \ve b \in D(A \ve x + \ve b)$, and so together with the
definition of the decodable noise region, we have
\begin{equation}
-\nabla d(A \ve x + \ve b, \ve r) \in N_{\cal Q}(A \ve x + \ve b).
\end{equation}
Since the function $d(\cdot, \cdot)$ is convex with respect to the first argument, 
we can use Lemma \ref{KKT} to verify that the output from the relaxed ML decoder is 
$\hat{\ve s} = A \ve x + \ve b$. Applying the inverse affine map, we can obtain the correct estimate:
\begin{equation}
\hat {\ve x} = A^{-1}(\hat{\ve s}-\ve b) = \ve x.
\end{equation}
\hfill\qed
\end{corollary}

The decodable noise region of $\ve x$ completely characterizes the 
block error probability when the vector $\ve x \in {\cal P}$ is sent.
We here consider the probability of the event that 
the estimate $\hat{\ve x}$ obtained from the relaxed ML decoder 
coincides with the transmitted codeword vertex $\ve x \in {\cal P}$.
This probability is expressed as 
expressed as 
\begin{eqnarray}
P_C(\ve x) 
&\defeq& {\rm Prob}\{\ve x = \hat{\ve x} \} \\
&=& {\rm Prob}\{\ve z \in D(A\ve x+\ve b) \} \\ \label{correctprob}
&=& \int_{D\left(A \ve x+ \ve b \right)} p(\ve z) dz_1 dz_2 \cdots dz_n,
\end{eqnarray}
where $p(\ve z)$ is the probability density function of the additive noise $\ve z$.

The next example considers the case of the Gaussian linear vector channel.
\begin{example}
Assume that $z_i$ is an independent Gaussian random variable 
with zero mean and variance $\sigma^2$, that is, assume that the noise PDF $p(\ve z)$ is given by
\begin{equation} \label{gaussianpdf}
p(\ve z) = 
\frac{1}{(2 \pi \sigma^2)^{n/2}} \exp 
\left(- \sum_{i \in [1,n]}\frac{z_i^2}{2 \sigma^2}  \right). 
\end{equation}
  For this case, 
$d(\ve s, \ve r) \defeq ||\ve r - \ve s||^2$ is used as the distance measure for 
the relaxed ML decoder.  It is easy to verify that 
\begin{equation}
-\nabla d(\ve s, \ve r)  \propto (\ve r - \ve s)
\end{equation}
holds. From this proportional relation, we can obtain the equivalence relation
\begin{equation} \label{eqivrelation}
D (\ve s) =  N_{\cal Q} (\ve s).
\end{equation}
Suppose that a codeword vertex $\ve x \in {\cal P}$ is sent and 
$\ve r = A \ve x + \ve b + \ve z$ is observed at the receiver side.
By using the equivalence relation (\ref{eqivrelation}) and substituting for $ p(\ve z) $ from equation (\ref{gaussianpdf}) 
into equation (\ref{correctprob}), we have the correct decision probability $P_C(\ve x)$ of 
the relaxed ML decoder for the Gaussian linear vector channel:
\begin{equation}
P_C(\ve x)  = 
\frac{1}{(2 \pi \sigma^2)^{n/2}} \int_{N_{\cal Q} (A\ve x + \ve b)} \exp 
\left(- \sum_{i \in [1,n]}\frac{z_i^2}{2 \sigma^2}  \right) d \ve z, 
\end{equation}
where $d \ve z \defeq dz_1 dz_2 \cdots dz_n$.
\hfill\qed
\end{example}
This example shows that the error performance of the relaxed ML decoder is
dominated by the shape of the set of normal cones $N_{\cal Q} (A\ve x + \ve b)$ where
$\ve x$ is a codeword vertex of the fundamental polytope ${\cal P}$.

\subsection{Analysis for successive decoding}
It is highly desirable to perform another decoding process, which is called {\em secondary decoding},
after a relaxed MLD process.  This is because the relaxed ML decoder may output
a non-vertex point as the solution of the minimization problem. If such a non-vertex point 
is close enough to the transmitted vertex,  it can be corrected with a secondary decoding process.
 Secondary decoding can thus improve the overall decoding performance.

The successive decoding process is given as follows.
\begin{eqnarray} \label{successive1st}
\hat{\ve x}_1 &=& \arg \min_{\ve x \in {\cal P}} d(A \ve x + \ve b, \ve r) \\ \label{successive2nd}
\hat{\ve x}_2 &=& \Gamma(\hat{\ve x}_1).
\end{eqnarray}
The decoding rule (\ref{successive1st}) is just the relaxed MLD rule, while
the rule (\ref{successive2nd}) corresponds to the secondary decoding.
The function $\Gamma: {\cal R}^n \rightarrow {\cal R}^n$ represents the decoding function for
secondary decoding. For example, $\Gamma \defeq (\Gamma_1,\Gamma_2,\ldots, \Gamma_n)$ 
which is defined by
\begin{equation}
\Gamma_i(a_i) = 
\left\{
\begin{array}{cc}
0, & a_i <  0.5  \\
1, & a_i \ge 0.5 \\
\end{array}
\right.
\end{equation}
is a possible secondary decoding function. This function quantizes a fractional value in the output 
vector from the relaxed ML decoder. Another example of the secondary decoding is the 
built-in min-sum decoder implemented in the interior point decoding presented in the next section.

In the following analysis, the {\em decision region} of the secondary decoding plays a crucial role.
The decision region of $\ve x$, where $\ve x$ is a codeword vertex of ${\cal P}$, is given by 
\begin{equation}
\Delta(\ve x) \defeq \{\ve r \in {\cal R}^n: \ve x = \Gamma(\ve r)  \}.
\end{equation}
We assume that $\Delta(\ve x)$ and $\Delta(\ve x')$ are disjoint if $\ve x \ne \ve x'$.
In the following, we assume a Gaussian linear vector channel for simplicity.

\begin{definition}[Union of the shifted normal cones]
The union of the shifted normal cones associated with the successive decoding 
is defined by
\begin{equation}
T(\ve x) \defeq \bigcup_{\ve x' \in \Delta(\ve x)}
\left[N_{\cal Q}(A \ve x' + \ve b) + (A \ve x' + \ve b) \right].
\end{equation}
\end{definition}

\begin{figure}[htbp]
\begin{center}
\includegraphics[scale=0.8]{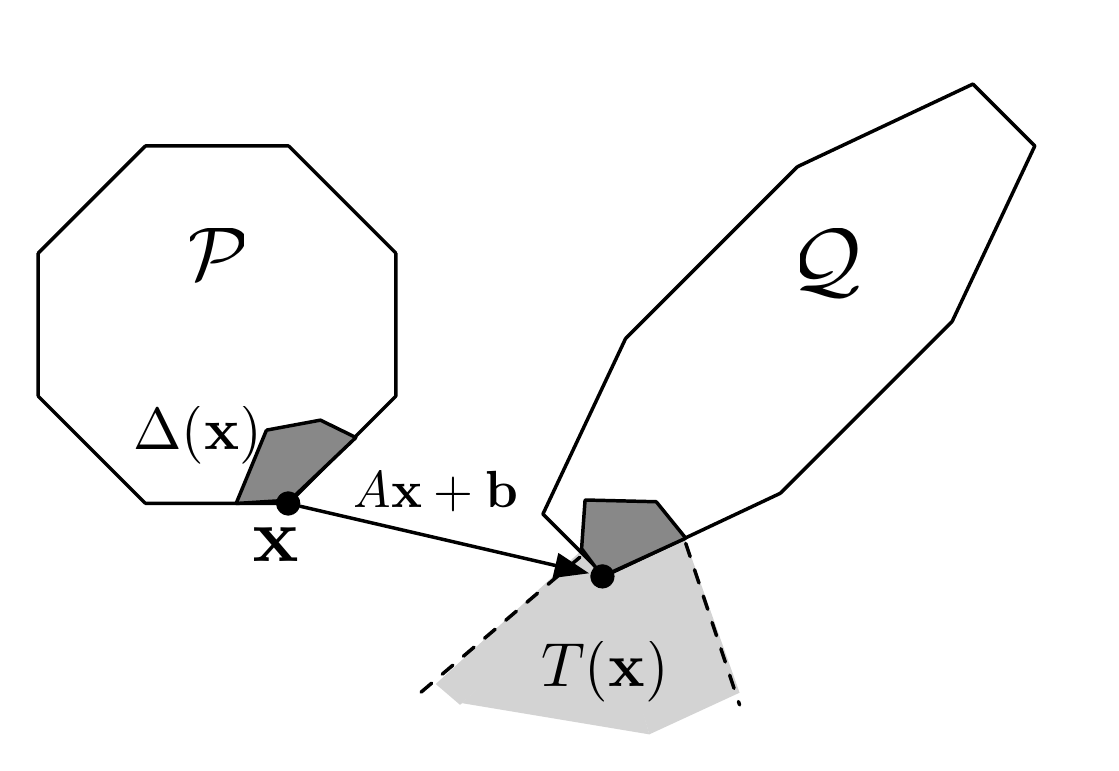} \\
  \caption{Decision region of the successive decoding.}
  \label{fig-decisionregion}
\end{center}
\end{figure}

The next corollary shows that the set $T(\ve x)$ can be considered to be the decision region corresponding to 
the transmitted  vector $\ve x$.
\begin{corollary}
\label{decisioncoro}
Assume that $\ve x \in {\cal P}$ is a codeword vertex and is sent to the channel.
The receiver obtains $\ve r = A \ve x + \ve b + \ve z$ as the channel output.
The successive decoder outputs the correct estimate $\hat{\ve x} = \ve x$
if $\ve r \in T(\ve x)$. \\
(Proof) The proof is almost same as the proof of Corollary \ref{vertexcol}, and so is omitted. \hfill\qed
\end{corollary}
Figure \ref{fig-decisionregion} illustrates the decision region of the successive decoding.
We can see that $T(\ve x)$ totally includes the shifted normal cone of $\ve x$.
This means that secondary decoding can improve the correct decision probability $P_C(\ve x)$
in the case of Gaussian linear vector channels.

Due to Corollary \ref{decisioncoro}, error analysis of the successive decoding can be divided  into 
several sub-problems: (i) analysis of the local structure of the fundamental polytope, (ii) analysis of the 
the decision region $\Delta(\ve x)$ and (iii) numerical evaluation of the error probability (via multi-dimensional integration
or bounds). Of course, these sub-problems are not easy to solve for long binary linear codes. However, 
the geometrical perspective established in this section will become a solid basis of further error analysis and
it helps us to understand the behavior of a sub-optimal relaxed ML decoder.

\section{Overview of Interior Point Decoding}

Although the relaxed MLD problem is easier to solve than the original MLD problem, it is still 
a computationally difficult problem and we must consider techniques to solve it efficiently.  
Since the relaxed MLD problem can be seen as a convex optimization problem, it is 
reasonable to apply conventional optimization techniques to this problem.
In this section, the interior point decoding algorithm for linear vector channels
will be presented. The proposed decoding algorithm is based on the idea of the interior point algorithm for
convex optimization \cite{Boyd}.

In this section, we first briefly review  the idea of the interior point algorithm based on a barrier function method, before describing the overall structure of the proposed decoding algorithm. 
 
The details of the sub-procedures required for the algorithm are explained in the subsequent
subsections.

\subsection{Barrier function method: a brief review}

The interior point algorithm describes a class of optimization algorithms 
 for solving LP and convex optimization problems.
Many sophisticated interior point algorithms have been developed   \cite{Boyd}, here we
 explain the simplest algorithm which is based on a {\em barrier function method}.

Let $f(\ve x) (\ve x \in {\cal R}^n)$ be a real-valued function to be minimized. Assume that 
$f(\ve x)$ is a convex function and the feasible region ${\cal F}$ is a convex subset of
${\cal R}^n$. 
A convex optimization problem is a minimization problem with a constraint 
of the form
\begin{equation}\label{convexpro}
\mbox{minimize  } f(\ve x), \mbox{ s.t. } \ve x \in {\cal F}.
\end{equation}

The key idea of the barrier function method is to convert the original convex optimization problem (\ref{convexpro}) 
into an unconstrained optimization problem by using a barrier function.
Let $B(\ve x)$  be a  {barrier function} which has the following properties:
(i) $B(\ve x)$ takes a finite real value if $\ve x \in {\cal F}^*$ where ${\cal F}^*$ is the interior set of
the feasible region ${\cal F}$; (ii) $B(\ve x) = \infty$ if $\ve x \notin {\cal F}^*$; (iii) $B(\ve x)$ is 
differentiable and convex. Combining the original objective function $f(\ve x)$ with 
the barrier function $B(\ve x)$,  we have a new convex objective function which is called a {\em merit function}
$
\psi(\ve x) \defeq t f(\ve x)  + B(\ve x).
$
The parameter $t$ is a positive real number called  a {\em scale parameter}.

Thus the barrier function method replaces the original problem (\ref{convexpro}) by the optimization 
  problem  
\begin{equation} \label{barrierprob}
\mbox{minimize  } \psi(\ve x), \mbox{ s.t. } \ve x \in {\cal R}^n .
\end{equation}

It is observed  that the constraints in (\ref{convexpro}) are absorbed into the 
barrier function and thus problem (\ref{barrierprob}) is an unconstrained minimization problem for a 
convex function $\psi(\ve x)$. In order to solve this problem, we can therefore   exploit an efficient numerical optimization algorithm
such as the gradient descent method or the Newton method.  

Of course, the unconstrained problem (\ref{barrierprob}) and the original problem (\ref{convexpro}) are not
Identical, and the optimal point $\ve x_1^*$ of problem (\ref{convexpro}) will not, in general, coincide with 
the optimal point $\ve x_2^*$ of (\ref{barrierprob}).  However, when $t \rightarrow \infty$, 
we can expect that $\ve x_2^* \rightarrow \ve x_1^*$ because the effect from the barrier function becomes
relatively small in this limit.   
On the other hand, smaller values of $t$ improve the rate of convergence of the solution.  
As $t$ becomes larger, the barrier function approaches a discontinuous function, which in general tends to retard the rate of convergence.  

\begin{figure}[htbp]
\begin{center}
\includegraphics[scale=0.3]{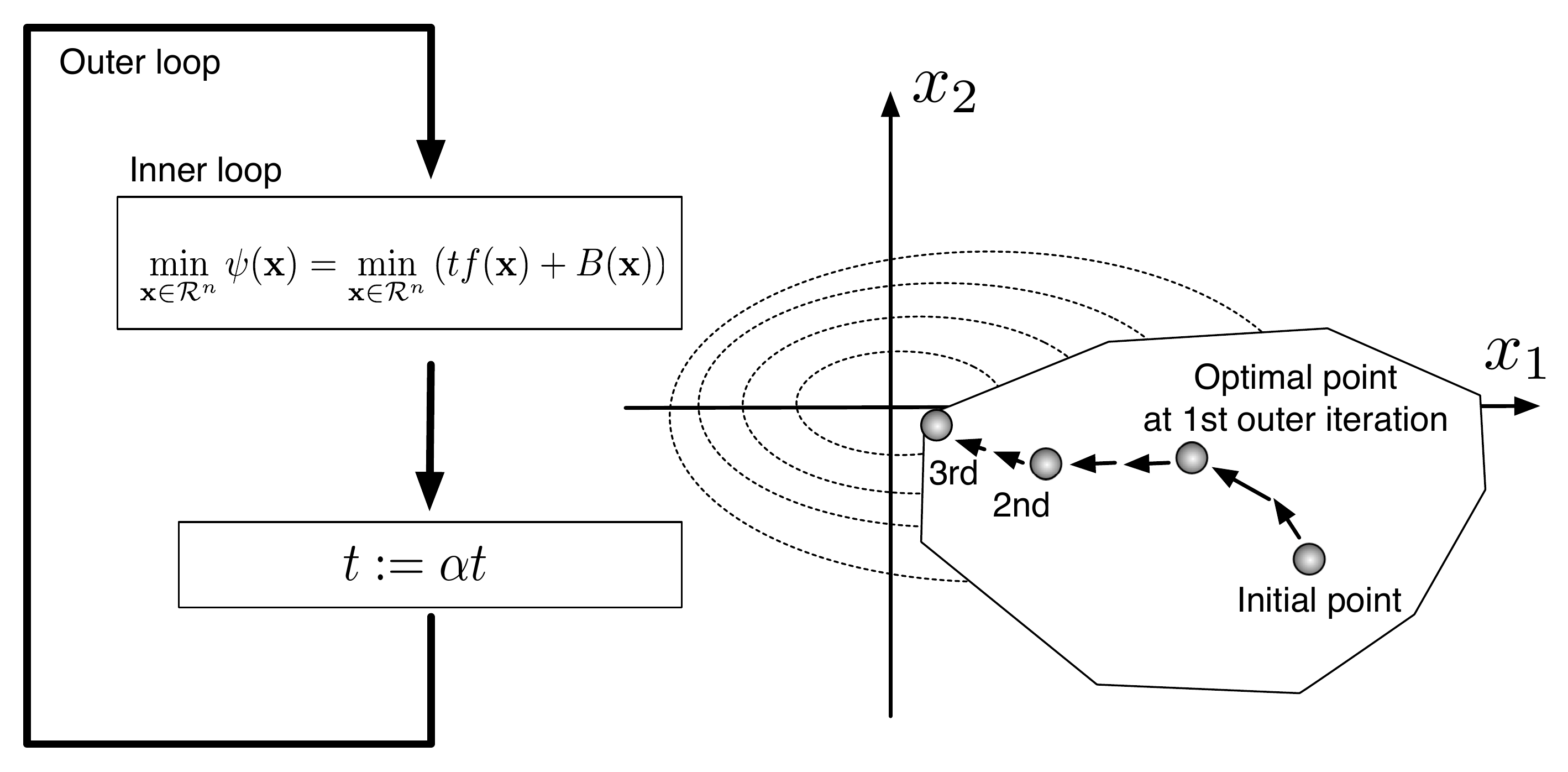} \\
  \caption{Interior point algorithm based on the barrier function method.}
  \label{diagram1}
\end{center}
\end{figure}

A well-known recipe to obtain a point $\ve x_2^*$ close to $\ve x_1^*$ is to 
perform the gradient descent or the Newton method several times
while $t$ is gradually increased. The interior point algorithm based on the barrier function method
consists of the two loops called {\em inner and outer loops}, respectively (see Fig.\ref{diagram1} left).
In the inner loop, $\psi(\ve x)$ is minimized using the gradient descent or the Newton method.
The aim of the outer loop is that of a scaling of $t$. In each iteration of the outer loop,  $t$ is multiplied by  
a positive constant $\alpha$, and so  the value of $t$ increases as the number of  outer iterations increases. 

The name "interior point" comes from the fact that
 search points (the tentative candidates for the optimal point) located on a trajectory moving towards the optimal point
are always contained in ${\cal F}^*$ (see Fig.\ref{diagram1} right).  For faster convergence, 
it is hoped that the trajectory of the search points does not approach 
the boundary of the feasible region. The barrier function is introduced to 
prevent a search point from approaching the boundary.

\subsection{Objective and merit functions for interior point decoding}

We here apply  the interior point algorithm  based on
the barrier function method to the relaxed MLD problem  (\ref{relaxedMLD}).
In the following, we will assume a Gaussian linear vector channel for simplicity, but the
extension to the non-Gaussian (or correlated Gaussian)  case is straightforward.
The objective function to be minimized is, therefore, $|| \ve r - (A \ve x + \ve b) ||^2$.
\begin{definition}[Objective function]
The objective function $f(\ve x) (\ve x \in {\cal R}^n)$ for the relaxed MLD problem is defined by
\begin{eqnarray}\nonumber
f(\ve x) &\defeq& || \ve r - (A \ve x + \ve b) ||^2 \\ \label{objectivegeneral}
&=& \sum_{i \in [1, n]} \left(r_i - \left(\sum_{j \in [1,n]} a_{ij} x_j + b_i \right) \right)^2,
\end{eqnarray}
where $r_i$ and $b_i$ are $i$th elements of $\ve r$ and $\ve b$, respectively. 
The symbol $a_{ij}$ denotes the $(i,j)$-element of the interference matrix $A$.
\hfill\qed
\end{definition}

The function $f(\ve x)$ is a differentiable, convex function, with these properties making it suitable    for use in the
the gradient descent and Newton methods. 
In the case of non-Gaussian linear vector channels,
we can also define an appropriate objective function. 

 Various choices exist for the form  of the barrier function.  In this paper, we adopt 
$b(x) \defeq -\ln(-x)$ as the basis of the barrier function, which is called a {\em log-barrier function}.
It is clear that  $b(x) =  \infty$ holds when $x = 0$;
on the other hand,  $b(x)$ takes a finite value if $x < 0$. Furthermore, $b(x)$ is a convex function.
Thus the function $b(x)$ satisfies the requirements
of the barrier functions as discussed above. Moreover, the derivative of $b(x)$, namely,
\begin{equation}
\frac{d}{dx} b(x) = -\frac{1}{x}
\end{equation}
is simple enough for  implementation in a decoding algorithm.
The log-barrier function including the parity constraints and box constraints is given below.
\begin{definition}[Log-barrier function]
The log-barrier function  for the fundamental polytope $B(\ve x)(\ve x \in {\cal R}^n)$ is defined by
\begin{eqnarray} \nonumber
B(\ve x) \hspace{-3mm}
&\defeq&  \hspace{-3mm}
- \sum_{i \in [1,m]} \sum_{S \subset T_i}
\ln \left[- \left(1 +\sum_{t \in S} (x_t-1) - \sum_{t \in A_i \backslash S} x_t \right) \right] \\
&-& \hspace{-3mm}\sum_{j \in [1,n]} \ln \left[- (- x_j) \right] - \sum_{j \in [1,n]} \ln \left[- (x_j-1) \right].
\end{eqnarray}
\hfill\qed
\end{definition}
The log-barrier function $B(\ve x)$ inherits the properties of $b(x)$;
$B(\ve x)$ is a convex and differentiable function. 
Furthermore, if $\ve x \notin {\cal P^*}$, then $B(\ve x) = \infty$ holds;
otherwise $B(\ve x) < \infty$.

The definition of the merit function including both the objective function and the log-barrier function 
is as follows.
\begin{definition}[Merit function]
For any $\ve x \in {\cal R}^n$, the merit function $\psi^{(t)}(\ve x)$ for 
the relaxed ML decoding problem is given by
\begin{equation}
\psi^{(t)}(\ve x) 
\defeq t f(\ve x) + B(\ve x),
\end{equation}
where $t$ is a positive real number.
\hfill\qed
\end{definition}
The first term of $\psi^{(t)}(\ve x)$ is a scaled version of 
the objective function.
The second term is the log-barrier function corresponding to the fundamental polytope.
Since the sum of convex functions is also a convex function, 
$\psi^{(t)}(\ve x)$ is a convex function. Note that $\psi^{(t)}(\ve x)$ takes a
finite value if $\ve x \in {\cal P}^*$.

\subsection{Partial Response Channels}

In the field of magnetic recording, the PR channel model is often exploited as a basis of system design.
Interferences arising from neighboring symbols are linearly superimposed on the current symbol, with these 
 interferences increasing the complexity of the decoding process.

Let $\delta+1$-real numbers $\{h_0,h_1, \ldots, h_{\delta} \}$ be the {\em partial response (PR) 
coefficients}. The parameter $\delta$ is called the {\em degree} of the PR channel.
The PR channel can be regarded as a discrete time finite impulse response (FIR) filtered channel
with additive white Gaussian noises (see Fig.\ref{prchannel2}). The definition of PR channel is given below.

\begin{definition}[Partial response channel]
A binary vector $\ve x =(x_1,x_2,\ldots, x_n) \in C$ is 
transmitted to the channel defined by 
\begin{equation}\label{prchannel}
r_j = \sum_{d=0}^\delta h_d (1 - 2 x_{j-d}) + z_j,\quad j \in [1,n], 
\end{equation}
where $z_j (j \in [1,n])$ is an independent Gaussian random variable with 
mean 0 and variance $\sigma^2$. 
By convention, we assume that
\begin{equation}
x_{-(\delta-1)} = x_{-(\delta-2)} = \cdots = x_{-1} = x_0 = \frac 1 2,
\end{equation}
which simplifies the treatment of the boundary condition.  
The channel is called a PR channel.
The signal to noise ratio ( $snr$ ) of the PR channel is defined by
\begin{equation}
snr \defeq \left(\sum_{d =0}^{\delta} h_d^2 \right)/ \sigma^2.
\end{equation}
\hfill\qed
\end{definition}
\begin{figure}[htbp]
\begin{center}
\includegraphics[scale=0.45]{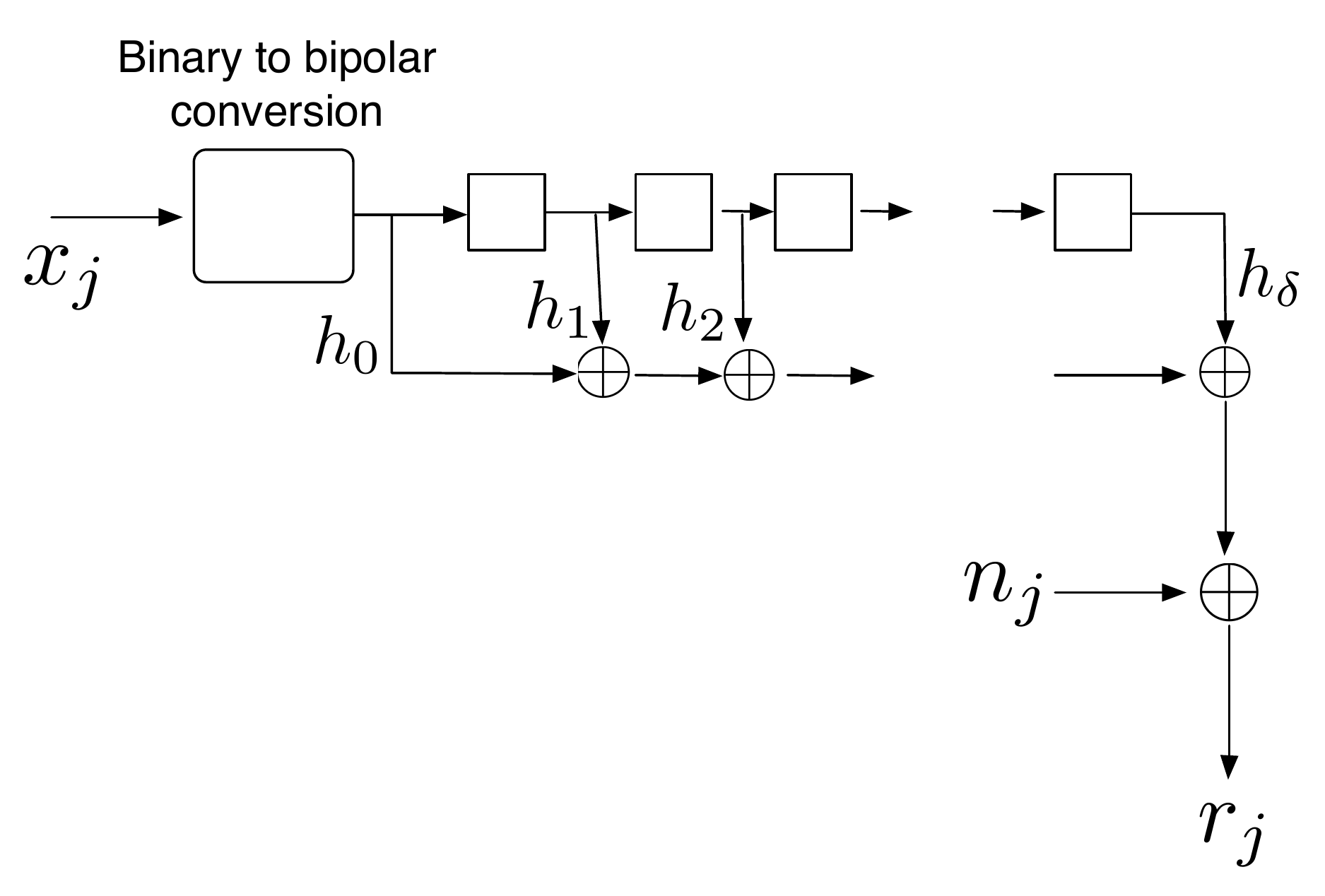} \\
  \caption{Block diagram of partial response channels.}
  \label{prchannel2}
\end{center}
\end{figure}

As shown in Fig.\ref{prchannel2}, the $j$th received symbol $r_j$ consists of
the weighted sum of  $\{1-2 x_j, 1-2x_{j-1},\ldots, 1-2x_{j-\delta} \}$ and the noise term.
It is clear that the PR channel can be expressed in the following linear vector channel form:
\begin{equation}
\ve r = A \ve x + \ve b + \ve z,
\end{equation}
where $A$ and $\ve b$ are given by 
\begin{equation}
A = 
-2 \left(
\begin{array}{cccc}
h_0 & \cdots & & \\
h_1 & h_0 & \cdots & \\
\vdots &\vdots & \vdots& \vdots\\
h_\delta & \cdots & h_0 &\cdots \\
\vdots &\vdots & \vdots& \vdots\\
\cdots& h_\delta & \cdots & h_0 
\end{array}
\right),
\ve b = \left(
\begin{array}{c}
h_0 \\
h_1 + h_0 \\
\vdots \\
\sum_{d=0}^{\delta-1} h_d \\
\sum_{d=0}^\delta h_d \\
\vdots \\
\sum_{d=0}^\delta h_d \\
\end{array}
\right).
\end{equation}
The PR channels can be regarded as  linear vector channels with a sparse interference matrix $A$.
Thus, the PR channel is an ideal candidate for the interior point decoding.
Throughout the paper, this channel will be  used as an example of a Gaussian linear vector channel.

For PR channels, the objective function $f(\ve x)$ takes the following form:
\begin{equation} \label{prcost}
f(\ve x) = \sum_{j = 1}^{n} \left(r_j - \left(\sum_{d=0}^\delta h_d (1 - 2 x_{j-d})  \right)   \right)^2.
\end{equation}
The partial derivative of $t f(\ve x)$ with respect  to a variable $x_p (p \in [1,n])$ is required to compute the approximate 
gradient presented later.  From the objective function (\ref{prcost}), we immediately have
\begin{eqnarray} \nonumber
\frac{\partial}{\partial x_p}t f(\ve x)\hspace{-3mm}
&=&\hspace{-3mm} t \sum_{j = 1}^{n} \frac{\partial}{\partial x_p}\left(r_j - \sum_{d=0}^\delta h_d (1 - 2 x_{j-d}) \right)^2 \\ %\nonumber
%&=& t \sum_{j = 1}^{n} 4 h_{j-p} \left(r_j - \left(\sum_{d=0}^\delta h_d (1 - 2 x_{j-d})  \right)   \right) I[ j-\delta \le  p \le j] \\ 
\label{prpartial}
&=& \hspace{-3mm}
4t \sum_{j = k}^{p+\delta}  h_{j-p} \left(r_j - \sum_{d=0}^\delta h_d (1 - 2 x_{j-d})    \right).
\end{eqnarray}
It can be observed that the number of additions required to evaluate equation (\ref{prpartial}) is proportional to  $\delta^2$.
This means that evaluation of the gradient $\nabla t f(\ve x)$ takes $O(n)$ computational time if
$\delta$ is constant.

We next consider the Hessian of the objective function, which is required for the Newton method to be discussed later.
From (\ref{prpartial}), we have 
\begin{eqnarray} \nonumber
\frac{\partial}{\partial x_p} t f(\ve x) 
&=& 4 t \sum_{j=p}^{p+\delta} h_{j-p} 
\left(r_j - \sum_{d=0}^\delta h_d (1 - 2 x_{j-d})   \right) \\
&=& 8 t \sum_{j=p}^{p+\delta}  \sum_{d=0}^\delta h_{j-p} h_d x_{j-d} + K
\end{eqnarray}
for $p \in [1,n]$. The symbol $K$ denotes terms that do not contain $x_k$.
Let $q \defeq j - d$. The second derivative of the objective function is given by 
\begin{eqnarray}
\frac{\partial}{\partial x_p x_q} t f(\ve x) 
&=& 8 t \sum_{j=p}^{p+\delta} h_{j-p} h_{j-q} I[j-q \in [0, \delta]].
\end{eqnarray}
By setting  $j - p = a$, we obtain the following expression: 
\begin{eqnarray}
\frac{\partial}{\partial x_p x_q} t f(\ve x) 
\hspace{-2mm}&=& \hspace{-2mm}8 t \sum_{a=0}^{\delta} h_{a} h_{a+p-q} I[a+p-q \in [0, \delta]]
\end{eqnarray}
for $p \in [1,n]$, $q \in [1,n]$.

\subsection{Overall structure of interior point decoding}

The goal of the interior point decoding is to solve the relaxed MLD problem by using 
an interior point algorithm based on the barrier function method.
The interior point decoding consists of two nested loops: an outer loop and an inner loop.
The inner loop corresponds to either an approximate gradient descent method or 
an approximate Newton method, both of which 
try to minimize the merit function $\psi^{(t)}(\ve x)$. The outer loop contains
the following sub-procedures:
inner loop, built-in min-sum decoding, and scaling of the parameter $t$.
The details of these sub-procedures are discussed in the following subsections.

The  procedure  $InteriorPoint(\cdot)$ is the main procedure for 
the interior point decoding.
The positive integer parameters $O_{max}$ and $I_{max}$ specify the number of
iterations executed in the outer  and inner loops, respectively.
The parameter $t_0$ is the initial scale parameter, which is a positive real number.

\vspace{0.3cm}
\fbox{
\begin{minipage}{8cm}
\small
\begin{description}
\item[Procedure: ] \hspace{5mm} $\hat{\ve b} := InteriorPoint(\ve r)$
\item[Input: ]   $\ve r \in {\cal R}^n$ : received word
\item[Output: ]  \ $\hat{\ve b} \in F_2^n \bigcup \{*\}$: estimation word
\item[Step 1 ] Let $\ve x := (1/2,1/2,\ldots, 1/2)$ and $t := t_0$.
\item[Step 2 ] Repeat the following sub-steps (2.1--2.4) $O_{max}$ times:
\begin{description}
\item[Step 2.1]\hspace{2mm} Repeat the following process $I_{max}$ times: 
\[
\ve x:=  InnerLoop(\ve x, t).
\]
\item[Step 2.2 ]\hspace{2mm} Execute $(p,\hat{\ve b}) := MinSum(\ve x)$.
\item[Step 2.3 ]\hspace{2mm} If $p = 0$, then exit.
\item[Step 2.4 ]\hspace{2mm} Let $t := \alpha t$.
\end{description}
\item[Step 3] Let $\hat{\ve b} := *$ ($*$ denotes decoding failure) and then exit.
\end{description}
\end{minipage}
}
\vspace{0.3cm}

The heart of the decoding algorithm is 
the process called $InnerLoop(\cdot)$. This procedure updates
a search point $\ve x$ using the gradient descent method or the Newton method so as to minimize the merit function $\psi^{(t)}(\ve x)$.
Since the search point $\ve x$ must always be contained in ${\cal P}^*$,
we need to check the feasibility of the search point in this process.

After the execution of the inner loop, 
{\em built-in min-sum decoding} is performed to obtain 
an estimate of the transmitted word.
The role of the min-sum decoding  is to find 
a codeword near to the current search point obtained from the inner loop process.

In the interior point method,  a search point must lie in the feasible region ${\cal P}^*$ in all iterations of the procedure.
The following lemma justifies the choice of  $\ve x = (1/2,1/2,\ldots, 1/2)$ as the
initial search point.
\begin{lemma}
If $H$ is a row-regular parity check matrix with $w_r \ge 3$, then
 the initial search point $\ve x = (1/2,1/2,\ldots, 1/2)$ is a feasible point, i.e.,
 $\ve x \in {\cal P}^*$.\\
 (Proof) It is evident that $\ve x$ satisfies the box constraints. Thus, we only need to 
 consider the parity constraints. For any $i \in [1,m], S \in T_i$,  we have
\begin{equation}
1 +\sum_{t \in S} \left(\frac 1 2 -1\right) - \sum_{t \in A_i \backslash S} \frac 1 2  = 1 - \frac 1 2  |A_i| = 1- \frac 1 2  w_r<0
\end{equation}
using the assumption $w_r \ge 3$ and row-regularity. This means that $\ve x$ also satisfies the parity constraints.  
\hfill\qed
 \end{lemma}

\subsection{Built-in min-sum decoding}

The advantage of exploiting the {\em built-in} min-sum decoding process after the inner loop process 
is the consequent reduction  in the required number of outer and inner iterations. 
If the current search point approaches close enough to a codeword vertex, 
the built-in decoding process can output the corresponding codeword as an estimate vector. In particular, we need not wait for the search point to converge to a  codeword vertex, which in general requires a 
much longer computational time.  
Furthermore, the built-in min-sum decoding acts as the secondary decoding discussed in 
the previous section\footnote{We can also use another decoding algorithm such as sum-product algorithm, 
bit-flipping algorithm instead of min-sum algorithm.}. That is, it can compensate a non-vertex point
obtained from an inner-loop process.
Therefore, this built-in decoding process is indispensable for the interior point decoding technique.

In the following, a brief description on the built-in min-sum decoding is given.
Let $B_j (j \in [1,n])$ be the set of row indices such that 
$
B_j \defeq \{i \in [1,m]: h_{ij} = 1 \}.
$
The following procedure $MinSum(\ve x)$ is the standard log-domain min-sum algorithm with
a {\em dumping factor} (also known as {\em normalized min-sum algorithm} \cite{Chen}).

\vspace{0.3cm}
\fbox{
\begin{minipage}{8cm}
\small
\begin{description}
\item [Procedure: ] \hspace{5mm}$(p, \hat{\ve b}) := MinSum(\ve x)$
\item [Input: ] $\ve x \in {\cal P}^*$: current search point 
\item[Output: ] \ $(p, \hat{\ve b})$: parity flag $p$ and tentative estimate vector $\hat{\ve b}$
\item [Step 1] Compute the log likelihood ratios:
\begin{equation}\label{LLR}
\lambda_j :=  \ln \left(\frac{1-x_j}{x_j} \right)
\end{equation}
for $j \in [1,n]$. Set $\xi_{i \rightarrow j} := 0$ for all pair $(i,j)$ satisfying $j \in A_i$.
\item [Step 2] Repeat the following sub-steps (2.1--2.4) $L_{max}$ times.
\begin{description}
\item [Step 2.1] \hspace{2mm} For all pairs $(i,j)$ satisfying $j \in A_i$, 
evaluate 
\begin{equation}
\eta_{j \rightarrow i} := \lambda_j + \sum_{k \in B_j \backslash i} \xi_{k \rightarrow j}.
\end{equation}
\item [Step 2.2 ] \hspace{3mm}For all pairs $(i,j)$ satisfying $j \in A_i$, 
evaluate
\begin{eqnarray} \nonumber
\xi_{i \rightarrow j} &:=& \kappa \left( \prod_{k \in A_i \backslash j}\mbox{sign}(\eta_{k \rightarrow i}) \right) \\ \label{coperation}
&\times& \hspace{-5mm}  \min_{k \in A_i \backslash j} |\eta_{k \rightarrow i}| .
\end{eqnarray}
The function $\mbox{sign}(\cdot)$ is defined by
\begin{equation}
\mbox{sign}(x) = 
\left\{
\begin{array}{cc}
1, & x \ge 0 \\
-1 & x < 0.  \\
\end{array}
\right.
\end{equation}
\item [Step 2.3 ] \hspace{2mm} For $j \in [1,n]$, decide the tentative estimate word $\hat{\ve b} = (\hat b_1,\hat b_2,\ldots,\hat b_n)$ 
in the following way:
\begin{equation}
\hat b_j := \left\{
\begin{array}{cc}
0, & \lambda_j + \sum_{k \in B_j} \xi_{k \rightarrow j} \ge 0\\
1, & \lambda_j + \sum_{k \in B_j} \xi_{k \rightarrow j} <  0.\\
\end{array}
\right.
\end{equation}
\item[Step 2.4 ] \hspace{3mm}If $H \hat{\ve b} = \ve 0$ holds, then let $p := 0$ and exit.
\end{description}
\item[Step 3] Let $p := 1$. 
\end{description}
\end{minipage}
}
\vspace{0.3cm}

In the initialization part, the LLRs used in the min-sum decoding  are
computed as expression (\ref{LLR}). Since $0 < x_j < 1$ holds for any $j \in [1,n]$, 
we here regard $x_j$ as the probability such that the $j$th transmitted symbol is 1.
The constant $\kappa$ which appears in expression (\ref{coperation}) is the dumping factor $(\kappa \simeq 0.7$--$0.9)$
which improves decoding performance of the min-sum decoding.

\section{Inner loop based on  gradient descent method}

The gradient descent method is a well-known minimization method for
an unconstrained convex function with a known  first derivative.
The convergence of the gradient descent method is relatively slow compared with methods
that utilize second derivatives (i.e., the Hessian) of the objective function, such as 
the Newton method, but the gradient descent method is much easier to implement.

\subsection{Brief review on gradient descent method}

We here briefly review the gradient descent method.
Let $h(\ve x), \ve x \in {\cal R}^n$ be a real-valued convex function to be minimized.
In the process of the gradient descent method, 
a search point  gradually approaches the optimal point. For each iteration of 
the minimization process, the search point $\ve x$ is updated as
\begin{equation}\label{gupdate}
\ve x := \ve x - s \nabla h(\ve x),
\end{equation}
where $\nabla h(\ve x)$ is the gradient of $h(\ve x)$.
That is, the search point moves in the descent direction $-\nabla h(\ve x)$.
A positive real parameter $s$ is called the {\em step size} parameter.

The optimal choice of $s$, for a given $\ve x$, is obtained by solving the following one-dimensional optimization problem:
\begin{equation}\label{exactlinesearch}
s = \arg \min_{s' > 0}h(\ve x - s' \nabla h(\ve x)).
\end{equation}
This one-dimensional optimization process is usually called the {\em exact line search}.
Other than  the exact line search (\ref{exactlinesearch}), some 
inexact line search methods such as the bisection scheme with backtracking \cite{Boyd} also exist.
For the exact line search and some of the inexact line searches, 
it can be proved that the search point eventually converges to 
the optimal point \cite{Boyd} using the gradient descent update (\ref{gupdate}).

\subsection{Approximate gradient descent method}

The procedure $InnerLoop(\cdot)$ returns a new search point which
is computed from the current search point. 
In order to make the interior point decoding computationally tractable, 
some approximations  are introduced.

The aim of the procedure $InnerLoop(\cdot)$ is to find
the minimum of $\psi^{(t)}(\ve x)$ using a gradient descent method.
It uses the approximate gradient 
(instead of the true gradient $\nabla \psi^{(t)}(\ve x)$) and
an inexact line search scheme inside. Thus, strictly speaking, 
the procedure is an approximation of the gradient descent method.

The following procedure is the main part of the inner loop, which  can be regarded as an 
inexact line search method based on the bisection scheme.

\vspace{0.3cm}
\fbox{
\begin{minipage}{8cm}
\small
\begin{description}
\item[Procedure: ] \hspace{5mm} $\ve x := InnerLoop(\ve x,t)$
\item[Input: ] $\ve x \in {\cal P}^*$: current search point
\item[Output: ] \ $\ve x$: updated search point
\item[Step 1 ] Set $s := 1$.
\item[Step 2 ]  $\ve g := ApproxGradient(\ve x, t)$.
\item[Step 3 ] Let
$
\tilde{ \ve   x} := \ve x - s \ve g
$
\item[Step 4]  If $IsFeasible(\tilde{\ve x}) = 0$, namely $\tilde{ \ve   x} \notin {\cal P}^*$ holds, 
then let $s := s/2$ and return to Step 3.
\item[Step 5] Let $\ve x := \tilde{ \ve   x}$. 
\end{description}
\end{minipage}
}
\vspace{0.3cm}

Since the vector $\ve x$ is an interior point of the fundamental polytope, 
there exits $s > 0$ satisfying  $\ve x - s \ve g \in {\cal P}^*$. 
This means that the loop composed of Step 3 and Step 4 must eventually stop.

\begin{figure}[htbp]
\begin{center}
\includegraphics[scale=0.4]{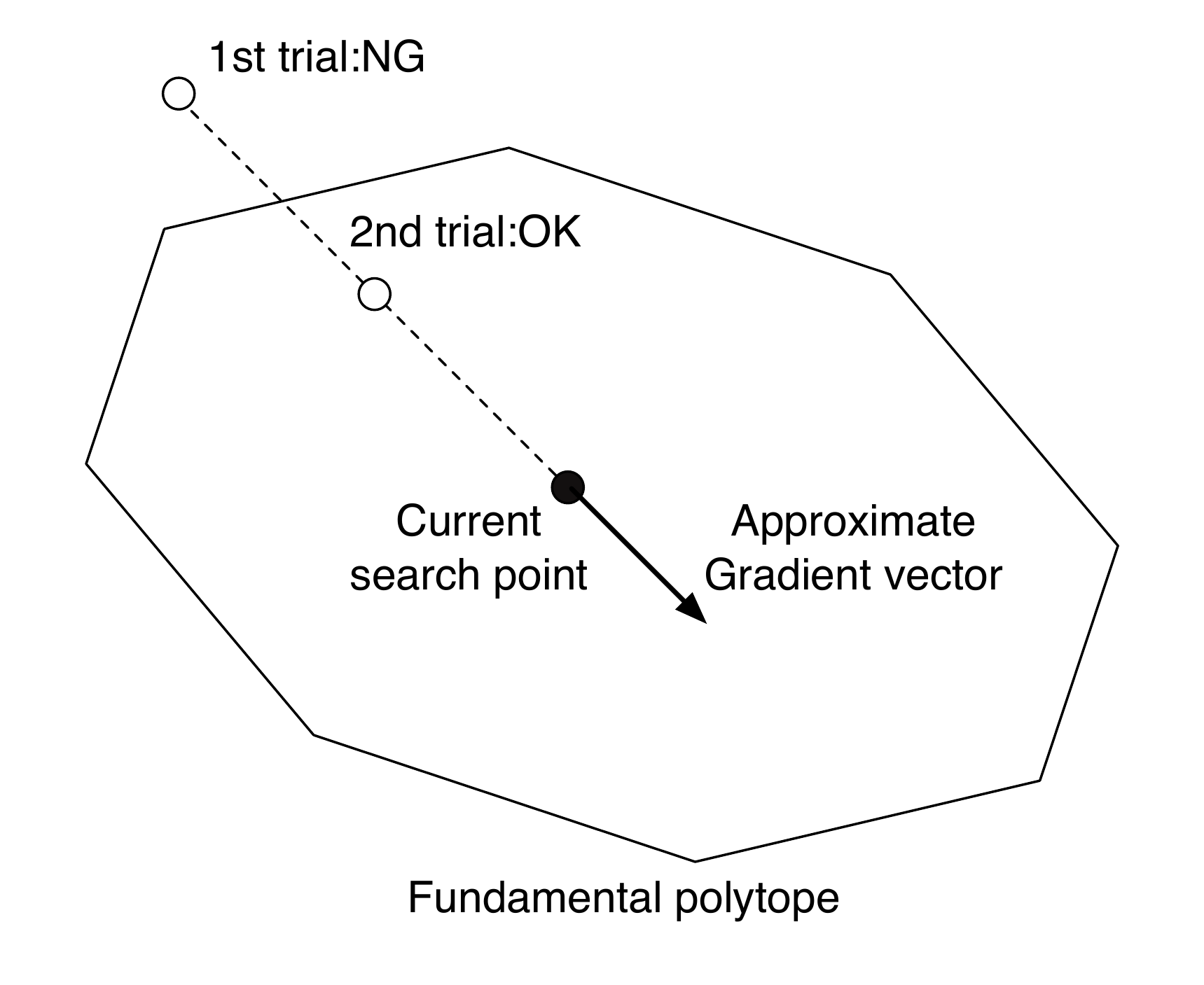} \\
  \caption{A search step of $InnerLoop(\cdot)$.}
  \label{aproxdescent}
\end{center}
\end{figure}

The process $InnerLoop(\cdot)$ includes the feasibility check in its procedure. Thus, it is guaranteed that 
the updated point  is located in the feasible region ${\cal P}^*$.
Figure \ref{aproxdescent} presents an example of a search step of $InnerLoop(\cdot)$. 
A search process is performed to find a feasible point on the half line starting from the current search point in the
opposite direction to that of the approximate gradient vector. 
For the case shown in Fig.\ref{aproxdescent}, the first temporary search point $\tilde{ \ve   x}$ is not in ${\cal P}^*$.
Thus, the step size parameter $s$ is reduced  to $s:=s/2$ to produce  the second 
temporary search point, which is a feasible point. 
As a result,  the second point becomes the next search point (i.e., the updated search point).

The procedure uses the approximate gradient $\ve g$ and does not include
an evaluation of the merit function $\psi^{(t)}(\ve x)$. Moreover, only a feasibility check is carried out in the procedure.
Thus, we cannot expect the objective function to be a non-increasing function of the number of iterations in this case.
This implies that  the search point may not converge to the optimal point.
However, due to this compromise, 
we obtain a huge reduction in computational costs. 
For example, the evaluation of the objective function,
which requires  a time computational complexity $O(2^{w_r} n)$,
can be avoided in this procedure. 
In this paper, we assume that $m$ scales proportional to $n$, namely $m  = (1-R) n$,  $0 < R < 1$.
In contrast, the execution of the feasibility check and evaluation of the approximate gradient
take only $O(w_r n)$ computation time.
Note that the evaluation of the true gradient of the barrier function $\nabla B(\ve x)$ requires $O(2^{w_r} n)$ computation time. 

\subsection{Feasibility check}

The purpose of the procedure $IsFeasible(\cdot)$  is to decide the feasibility of 
a given $\ve x \in {\cal R}^n$.  In the interior point method, a search point must lie in the 
feasible region in all iterations. Thus, the feasibility check is one of the key procedures
of the inner loop.

The procedure $IsFeasible(\ve x)$ returns 
1 if $\ve x \in {\cal P}^*$; otherwise it returns 0.  Of course, we could evaluate 
all the inequalities (\ref{nocodeword2}) and (\ref{box2}) to decide the feasibility of given $\ve x$. 
However, such a  brute-force approach 
takes $O(2^{w_r} n)$ computation time to check the feasibility
because the  number of inequalities (\ref{nocodeword2}) is $2^{w_r-1} m$. 
However, we can do much better as shown in the next lemma and the algorithm.

\begin{lemma}[Feasiblity check]
\label{lemmafeasiblecheck}
The point $(x_1,x_2,\ldots, x_n)\in {\cal R}^n$ belongs to ${\cal P}^*$
if and only if $\theta_i < 0$ for any $i \in [1,m]$ and $0 <  x_j < 1$ for any $j \in [1,n]$
where $\theta_i (i \in [1,m])$ is defined by
\begin{equation}
\theta_i \defeq \max_{S \subset T_i} \left[1 +\sum_{l \in S} (x_l-1) - \sum_{l \in A_i \backslash S} x_l  \right].
\end{equation}
(Proof)
If $\theta_i < 0$ for any $i \in [1,m]$ and  $0 <  x_j < 1$ for any $j \in [1,n]$,  then 
\begin{equation}
1 +\sum_{l \in S} (x_l-1) - \sum_{l \in A_i \backslash S} x_l <0
\end{equation}
holds for any $i \in [1,m]$ and $S \subset T_i$,  because $\theta_i$ is 
the maximum value of $1 +\sum_{l \in S} (x_l-1) - \sum_{l \in A_i \backslash S} x_l$.
This implies that $\ve x$ is a feasible point.  

We then consider the opposite direction. Assume that  $\ve x \notin {\cal P}^*$.
This means that some of the parity constraints or the box constraints are violated. 
If some of the box constraints are not satisfied, then there exists $j^* \in [1,n]$ such that
$x_{j^*}$ is not in the open interval between $0$ and $1$.
If some of the parity constraints are not satisfied, 
there exists an index $i^* \in [1,m]$ and $S^* \subset T_{i^*}$ such that
\begin{equation}
1 +\sum_{l \in S^*} (x_l-1) - \sum_{l \in A_{i^*} \backslash S^*} x_l  \ge 0.
\end{equation}
From the definition of $\theta_i$, it is evident that $\theta_{i^*} \ge 0$ in such a case.
This completes the proof. \hfill\qed
\end{lemma}

Lemma \ref{lemmafeasiblecheck} is useful to reduce the computational time required to carry out the 
feasibility check, because the
computational time required to evaluate  all $\theta_i (i \in [1,m])$ is proportional to 
$w_r m$. In other words, $\theta_i$ can be computed without generating 
all the subsets of  odd size  in $T_i$. We can use an idea  
similar to maximum likelihood decoding for even weight codes\footnote{The Viterbi algorithm can also be used for evaluating $\theta_i$. }.
The following procedure  includes such an idea for an
efficient computation of $\theta_i$.

\vspace{0.3cm}
\fbox{
\begin{minipage}{8cm}
\small
\begin{description}
\item[Procedure: ] \hspace{5mm} $f := IsFeasible(\ve x)$
\item[Input: ] $\ve x = (x_1,x_2,\ldots,x_n) \in {\cal R}^n$: current search point
\item[Output: ] \ $f$: If $\ve x \in {\cal P}^*,  $then $f := 1$; otherwise $f := 0$.
\item[Step 1] If there exists $j \in [1,n]$ such that $x_j$ do not satisfy $0 < x_j < 1$, 
set $f := 0$ and exit.
\item [Step 2] Repeat sub-steps (2.1--2.3) from $i := 1$ to $m$.
\begin{description}
\item[Step 2.1] \hspace{3mm} For $l \in A_i$, set
\begin{eqnarray}
\beta_l^{(i)} &:= &
\left\{
\begin{array}{cc}
0, & x_l -1 \le -x_l \\
1, & x_l -1 >  -x_l \\
\end{array}
\right. 
\end{eqnarray}
and evaluate
\begin{equation}\label{ueq}
u := 1 + \sum_{l \in A_i} \max\{x_l-1,-x_l \}, 
\end{equation}
and 
$
v := \sum_{l \in A_i} \beta_l^{(i)}.
$
\item[Step 2.2] \hspace{3mm} If $v$ is even, then 
update $u$ as
$
u := u - \min_{l \in A_i} |2 x_l -1|
$
and let 
$
\beta_{l_{min}}^{(i)} := \beta_{l_{min}}^{(i)} \oplus 1,
$
where 
$
l_{min} := \arg \min_{l \in A_i} |2 x_l -1|.
$
\item[Step 2.3] \hspace{3mm}If $u \ge 0$, then set $f := 0$ and exit.
\end{description}
\item[Step 3]  Set $f := 1$ and exit. 
\end{description}
\end{minipage}
}
\vspace{0.3cm}

The symbol $\oplus$ denotes addition over $F_2$.
Let us consider the states of the variables after an execution of the above procedure.
Let $S' \defeq \{l \in A_i:  \beta^{(i)}_l = 1\}$. Suppose the case where $v$ is odd.
From  this assumption, it is evident  that $|S'|$ is odd.  The right-hand side of (\ref{ueq}) can be 
rewritten  as
\begin{eqnarray} \nonumber
u &=& 
1 + \sum_{l \in A_i} \max\{x_l-1,-x_l \}  \\ \nonumber
&= &
1+ \sum_{l \in S'} (x_l-1) - \sum_{l \in A_i \backslash S'} x_l    \\ \nonumber
&= &
1+ \max_{S \subset T_i} \left[\sum_{l \in S} (x_l-1) - \sum_{l \in A_i \backslash S} x_l  \right] \\
&= & \theta_i.
\end{eqnarray}
Thus, in this case, $u = \theta_i$ holds. 
We then consider the case where $v$ is even. In such a case, one of the components in $\{ \beta_l^{(i)} \}_{l \in A_i}$ 
should be flipped so as to make the weight of the binary vector $\{ \beta_l^{(i)} \}_{l \in A_i}$ odd. 
Let $l'$ denote the index  at which the bit flip occurs, namely $\beta_{l'}^{(i)} := \beta_{l'}^{(i)} \oplus 1$.
The bit flipping decreases the value of $u$ to $u - |x_l -1- (-x_l) | = u - |2 x_l -1|$.
Since the aim of the bit flipping is to find the optimal subset with odd size, it is reasonable to 
determine the index $\l'$ according to the values of $ |2 x_l -1|$. Namely,
$t_{min} := \arg \min_{l \in A_i} |2 x_l -1|$ gives the smallest decrement (i.e., the largest value of $u$).
Therefore,   $u = \theta_i$ also holds for this case. 

\subsection{Approximate gradient}

The gradient of $\psi^{(t)}(\ve x)$, which is defined on ${\cal P}^*$, is given by
\begin{equation}
\nabla \psi^{(t)}(\ve x) \defeq \left(\frac{\partial}{\partial x_1} \psi^{(t)}(\ve x),
\frac{\partial}{\partial x_2} \psi^{(t)}(\ve x), \ldots, \frac{\partial}{\partial x_n} \psi^{(t)}(\ve x)  \right).
\end{equation}
We have, after some manipulation, 
the partial derivative of $\psi^{(t)}(\ve x)$ with respect to the variable $x_k (k \in [1,n])$: 
\begin{eqnarray} \nonumber
\frac{\partial}{\partial x_k} \psi^{(t)}(\ve x) \hspace{-2mm} &=& \hspace{-2mm} t \frac{\partial}{\partial x_k} f(\ve x) 
+\frac{\partial}{\partial x_k} B(\ve x) \\  \nonumber
&=&\hspace{-2mm} 
t \frac{\partial}{\partial x_k} f(\ve x) 
+\sum_{i \in [1,m]} \sum_{S \subset T_i}\tau_k^{(i,S)}(\ve x) \\
&-& \frac{1}{x_k} - \frac{1}{x_k - 1 },
\end{eqnarray}
where
\begin{equation}
\tau_k^{(i,S)}(\ve x) \defeq 
\frac{I[k \in A_i \backslash S]-I[k \in S] }{1+\sum_{l \in S}(x_l-1)-\sum_{l \in A_i \backslash S}x_l}
\end{equation}
for $i \in [1,m]$, $S \subset [1,n]$.
The notation $I[condition]$ is the indicator function
such that $I[condition] = 1$ if $condition$ is true; otherwise, it gives 0.
Note that the derivative of the objective function (\ref{objectivegeneral}) is given by
\begin{equation}
t \frac{\partial}{\partial x_k} f(\ve x) = 
-t\sum_{i \in [1, \ell]} 2 a_{ik} \left(r_i - \left(b_i +\sum_{j \in [1,n]} a_{ij} x_j  \right) \right).
\end{equation}
The next example presents the gradient of the merit function of PR channels.
\begin{example}
For  the case of PR channels,  we have the following derivative of the merit function:
\begin{eqnarray} \nonumber
\frac{\partial}{\partial x_k} \psi^{(t)}(\ve x) \hspace{-2mm}&=& 
4t \sum_{j = k}^{k+\delta}  h_{j-k} \left(r_j - \left(\sum_{d=0}^\delta h_d (1 - 2 x_{j-d})  \right)   \right) \\ \nonumber
&+& \sum_{i \in [1,m]} \sum_{S \subset T_i}\tau_k^{(i,S)}(\ve x) 
- \frac{1}{x_k} - \frac{1}{x_k - 1 }.
\end{eqnarray}
\hfill\qed
\end{example}

We can see that the partial derivative 
corresponding to the barrier function of the parity constraints contains $2^{w_r-1} m$ terms.
The computational time of computing $\nabla\psi^{(t)}(\ve x)$ is therefore an exponential function 
of the row weight $w_r$.
The computational cost of the gradient becomes the major obstacle to achieving a fast implementation of 
the interior point decoding when a given parity check matrix  has a relatively large row weight.
It is for this reason that we use an approximate gradient instead of the true gradient $\nabla  \psi^{(t)}(\ve x)$.

\begin{definition}[Approximate gradient]
The approximate gradient, denoted by $\ve g \defeq (g_1, \ldots, g_n) $,  
is defined by
\begin{equation}
g_k \defeq 
t \frac{\partial}{\partial x_k} f(\ve x) +\sum_{i \in [1,m]}\tau_k^{(i,S^{(i)})}(\ve x)  - \frac{1}{x_k} - \frac{1}{x_k - 1 }
\end{equation}
for $k \in [1,n]$ and  $\ve x \in {\cal P}^*$.
The subset $S^{(i)}$ is defined by
\begin{equation} \label{Sidef}
S^{(i)} \defeq 
\arg \max_{S \subset T_i} \left[1+ \sum_{l \in S} (x_l-1) - \sum_{l \in A_i \backslash S} x_l \right], \quad i \in [1,m].
\end{equation}
\hfill\qed
\end{definition}

From the definition of $S^{(i)}$,  it is clear that $S^{(i)}$ can be expressed as 
$
S^{(i)} = \{l \in A_i: \beta_l^{(i)} = 1\}
$
where $\{\beta_l^{(i)} \}_{l \in A_i}$ is the vector obtained after an execution of
$IsFeasible(\ve x)$. 
This means that evaluation of $\sum_{i \in [1,m]}\tau_k^{(i,S^{(i)})}(\ve x) $
requires a computational time proportional to $w_r m$, 
which is much faster than the computational time required for the
evaluation of $\sum_{i \in [1,m]} \sum_{S \subset T_i}\tau_k^{(i,S)}(\ve x) $, 
and so following the execution of $IsFeasible(\ve x)$, 
we can immediately evaluate the approximate gradient $\ve g$ using $\beta_l^{(i)}$.

The approximate gradient is defined on $\ve x \in {\cal P}^*$.  This implies 
\[
1+\sum_{l \in S^{(i)}} (x_l-1)- \sum_{l \in A_i \backslash S^{(i)}} x_l < 0
\]
holds for any $i \in [1,m]$. From this inequality and the definition of $S^{(i)}$, it is easy to show that
\begin{equation}
\left| \tau_k^{(i,S^{(i)})}(\ve x)  \right|  \ge \left|\tau_k^{(i,S)}(\ve x)  \right|  
\end{equation}
holds for any $k \in [1,n], i \in [1,m]$ and $S \subset T_i (S \ne S^{(i)})$.
This inequality indicates that the approximate gradient includes 
the largest (in terms of absolute values) contribution for each $i \in [1,m]$.

The following procedure $ApproxGradient(\cdot,\cdot)$ efficiently evaluates 
the approximate gradient $\ve g$.

\vspace{0.3cm}
\fbox{
\begin{minipage}{8cm}
\small
\begin{description}
\item[Procedure: ] \hspace{6mm}$\ve g := ApproxGradient(\ve x,t )$
\item[Input: ] $\ve x \in {\cal P}^*$: current search point
\item[Output:  ] \hspace{3mm}$\ve g$: approximate gradient
\item[Step 1] Set 
\begin{equation}
g_k := 
t \frac{\partial}{\partial x_k} f(\ve x) 
- \frac{1}{x_k} - \frac{1}{x_k - 1 }
\end{equation}
for $k \in [1,n]$.
\item[Step 2] Execute $IsFeasible(\ve x)$ to obtain $S^{(i)}, i \in [1,m]$.
\item[Step 3] Repeat sub-step 3.1 from $i := 1$ to $m$.
\begin{description}
\item[Step 3.1] \hspace{3mm}Let
\begin{equation}
g_k := g_k + \tau_k^{(i,S^{(i)})}(\ve x)
\end{equation}
for $k \in A_i$. 
\end{description}
\end{description}
\end{minipage}
}
\vspace{0.3cm}

The time complexity of each step is as follows.
If the interference matrix $A$ contains only $O(n)$ non-zero elements (i.e., $A$ is a sparse matrix), 
the initialization process (Step 1) takes $O(n)$-time.
As discussed before, Step 2 requires $O(w_r n)$-time. 
Finally, Step 3 needs $O(w_r n)$-time. In total, the evaluation of the approximate gradient
takes $O(w_r n)$-time.

\section{Inner loop based on the Newton method}

The Newton method is a numerical minimization algorithm with faster convergence than 
that of the gradient descent method.
The convergence speed of the Newton method is known to be quadratic around the optimal point.  It is thus appropriate to 
consider the Newton method for use as an optimization engine in the interior point decoding in order
to achieve better decoding performance.
However, we should be careful about the time complexity required for the execution 
of the Newton method, in which we need to handle the Hessian of the merit function.
In general, evaluation of the Hessian of the merit function takes $O(n^2)$-time, and
solving the {\em Newton equation} needs $O(n^3)$-time. The Newton equation is 
a linear equation 
$
G \ve d = - \ve g,
$
where $G$ is the Hessian ($n \times n$ real matrix) at the current point and
$\ve g$ is the gradient vector. The vector $\ve d$ is called the {\em  Newton step}.
Thus, it is important to fully utilize special structures of the merit function $\psi^{(t)}(\ve x)$; 
for instance,  sparseness of the Hessian.

\subsection{Inner loop based on the Newton method}
The inner loop  using the Newton method is shown below.
Most processes are identical with the inner loop based on the gradient descent method.
The differences are in Steps 2--4; the evaluation of the {\em approximate Hessian} (Step 2), derivation of  
the Newton step (Step 3), and update of the temporary search point (Step 4).
The details of the new processes introduced  here are presented in the subsequent subsections.

\vspace{0.3cm}
\fbox{
\begin{minipage}{8cm}
\small
\begin{description}
\item[Procedure: ] \hspace{6mm}$\ve x := InnerLoop(\ve x,t)$
\item[Input:] $\ve x \in {\cal P}^*$: current search point
\item[Output:] \hspace{3mm} $\ve x$: updated search point
\item[Step 1] Set $s := 1$.
\item [Step 2]  Evaluate 
\begin{eqnarray} \nonumber
\ve g &:=& ApproxGradient(\ve x, t), \\ \nonumber
G &:=&   ApproxHessian(\ve x, t).
\end{eqnarray}
\item [Step 3] Solve the Newton equation 
$
G \ve d = -\ve g.
$
\item[Step 4] Let
$
\tilde{ \ve   x} := \ve x - s \ve d
$
\item[Step 5] If $IsFeasible(\tilde{\ve x}) = 0$, that is if $\tilde{ \ve   x} \notin {\cal P}^*$ holds, 
then let $s := s/2$ and return to Step 3.
\item[Step 6] Let $\ve x := \tilde{ \ve   x}$. 
\end{description}
\end{minipage}
}
\vspace{0.3cm}

\subsection{Approximate Hessian}

The second derivative of the barrier function $B(\ve x)$ is given by
\begin{eqnarray} \nonumber
\frac{\partial}{\partial x_p x_q} B(\ve x) 
&=& \sum_{i \in [1,m]} \sum_{S \subset T_i} 
\tau_p^{(i,S)}(\ve x) \tau_q^{(i,S)}(\ve x) \\ \label{hessianbx}
&+&I[p = q] \left(\frac{1}{x_p^2}+\frac{1}{(x_p-1)^2} \right) 
\end{eqnarray}
for $p \in [1,n]$, $q \in [1,n]$.
As in the case of the gradient descent method, a straightforward evaluation of
$(\partial B(\ve x))/(\partial x_p x_q) $ needs $O(2^{w_r} n)$-time. This is one of the reasons why
we will introduce the approximate Hessian given below.

\begin{definition}[Approximate Hessian of $\psi^{(t)}(\ve x)$]
\label{aproxhessian}
The approximate Hessian $G \defeq \{G_{pq} \}(p \in [1,n]$, $q \in [1,n]) $
is defined by
\begin{eqnarray} \nonumber
G_{pq} &\defeq& t \frac{\partial}{\partial x_p x_q} f(\ve x)  \\ \nonumber
&+& I[p = q]  \sum_{i \in [1,m]}
\tau_p^{(i, S^{(i)})}(\ve x) \tau_q^{(i, S^{(i)})}(\ve x)  \\
&+& I[p = q]\left(\frac{1}{x_p^2}+\frac{1}{(x_p-1)^2}  \right) 
\end{eqnarray}
for $p \in [1,n]$, $q \in [1,n]$. The subsets $S^{(i)} (i \in [1,m])$ are defined by (\ref{Sidef}). \hfill\qed
\end{definition}

We can see that the approximate Hessian includes a contribution from the (true) Hessian of the 
objective function $f(\ve x)$ and contribution from the {\em approximate Hessian of the
barrier function} $B(\ve x)$. The approximate Hessian of $B(\ve x)$ has only diagonal 
elements; non-diagonal elements are discarded. Another approximation used here is that 
the double summation $\sum_{i \in [1,m]} \sum_{S \subset T_i}$ in the left-hand side of equation (\ref{hessianbx})
is replaced with a single summation $\sum_{i \in [1,m]}$ by using $S^{(i)}$.
This approximation has already been used to derive the approximate gradient.
Due to these approximations, computation of $G$ requires only $O(w_r n)$-time if 
the interference matrix is sparse, i.e., the number of non-zero elements in $A$ scales as $O(n)$.

The process $ApproxHessian(\cdot, \cdot)$ is  the  routine to evaluate the approximate Hessian $G$
according to Definition \ref{aproxhessian}. Thus, the details are omitted.

The following example treats the case of the PR channel.
\begin{example}
For PR channel case, the approximate Hessian $G = \{G_{pq}\}$ has the following form:
\begin{eqnarray} \nonumber
G_{pq} &\defeq& 
8 t \sum_{a=0}^{\delta} h_{a} h_{a+p-q} I[a+p-q \in [0, \delta]]  \\ \nonumber
&+&  I[p = q] \sum_{i \in [1,m]}
\tau_p^{(i, S^{(i)})}(\ve x) \tau_q^{(i, S^{(i)})}(\ve x)  \\ 
&+& I[p = q] \left(\frac{1}{x_p^2}+\frac{1}{(x_p-1)^2} \right).
\end{eqnarray}
Figure \ref{aproxHessian} illustrates the configuration of the approximate Hessian.
In this case, the matrix $G$ becomes a symmetric Toeplitz matrix. Moreover, it has 
diagonal band structure as shown in the figure.   \hfill\qed
\begin{figure}[htbp]
\begin{center}
\includegraphics[scale=0.45]{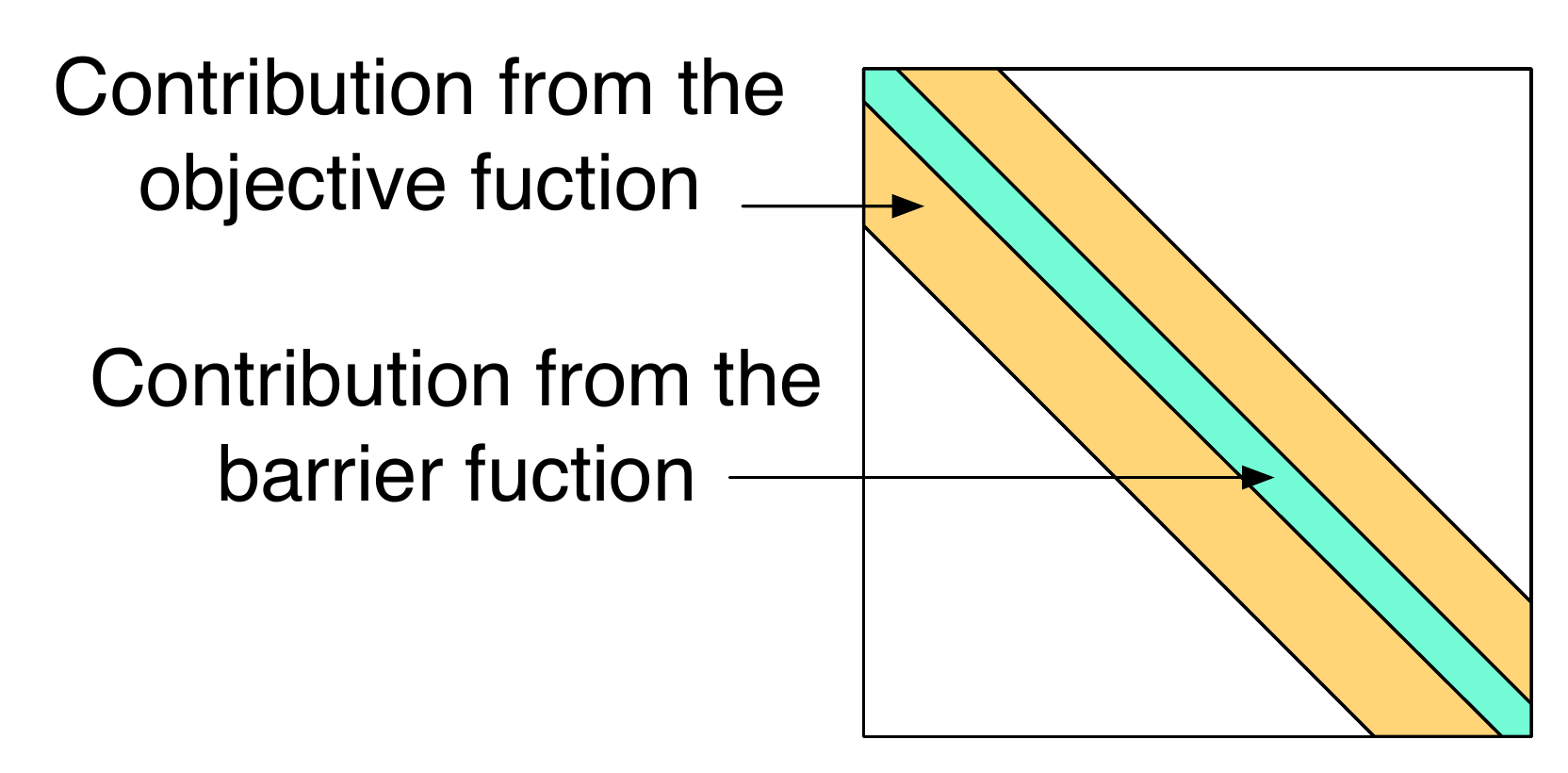} \\
  \caption{Configuration of the approximate Hessian for PR channels.}
  \label{aproxHessian}
\end{center}
\end{figure}
\end{example}

\subsection{Solving the Newton equation}

The most time consuming part of the Newton method is the determination of the Newton step, which is required to solve the Newton equation $G \ve d = - \ve g$. 
We here discuss how to solve the Newton equation efficiently in an inner-loop process.

\subsubsection{Cholesky decomposition}
If the approximate Hessian is a positive definite symmetric matrix,
Cholesky decomposition is applicable to solve the Newton equation.
For a given $n \times n$ positive definite matrix $M$, 
Cholosky decomposition decomposes $M$ as $M = L L^t$ where $L$ is a lower triangular matrix.
A linear equation $M \ve x = \ve b$ can be solved in an efficient way by using a combination of  
Cholesky decomposition and backward/forward substitutions \cite{Numerical}. 

In the case of PR channels,  the approximate Hessian becomes a symmetric 
positive definite matrix, and so we can employ Cholesky decomposition to solve the Newton equation.
Fortunately, Cholesky decomposition can be accomplished with time complexity
$O(n)$ for a matrix  having the form shown in Fig.\ref{aproxHessian}.
Since the backward/forward substitutions require only $O(n)$-time, the Newton equation
can be solved with time complexity $O(n)$.
This approach may be suitable for software implementation of the interior point decoding 
since Cholesky decomposition is a serial-type algorithm.

\subsubsection{Jacobi method}

The Jacobi method is an iterative method for solving a linear equation $M \ve x = \ve b$ \cite{Numerical}.
This method is especially suitable for the case where the coefficient matrix $M$ is sparse. 
The details of the Jacobi method are as follows.
The linear equation $M \ve x = \ve b$ can be rewritten in the following form:
\begin{equation}\label{jacobieq}
(L+U+D) \ve x = \ve b, 
\end{equation}
where $L$, $U$ and $D$ are lower triangular, upper triangular and diagonal matrices, respectively.
 Equation (\ref{jacobieq}) can be transformed  to
\begin{equation} \label{update}
\ve x = D^{-1} (\ve b - (L+U) \ve x ).
\end{equation}
We can regard equation (\ref{update}) as an update rule of $\ve x$ and  thus obtain the following recursive formula:
\begin{equation} 
\ve x^{(k)} = D^{-1} (\ve b - (L+U) \ve x^{(k-1)} ).
\end{equation}
Starting from an appropriate initial value $\ve x^{(0)}$,  
we can evaluate the above  recursive formula  iteratively. 
If certain conditions are met, $\ve x^{(k)}$ eventually converges to the solution of the linear equation 
$M \ve x = \ve b$.

Sometimes the Jacobi method fails to converge when the matrix $M$ is not diagonally dominant.
Under relaxation (UR)\footnote{UR is known as Successive Over Relaxation (SOR) when $1 < \omega< 2$.} 
of the Jacobi method yields convergent results 
for a wider class of linear equations, including those that cannot be treated with by the original Jacobi method.
The update rule for the UR Jacobi method is given by  
\begin{eqnarray}
\ve x^{(k)} &=& \omega \ve y^{(k)} + (1-\omega)\ve x^{(k-1)} \\
\ve y^{(k)}  &=& D^{-1} (\ve b - (L+U) \ve x^{(k-1)} ).
\end{eqnarray}
where $\omega$ is a real constant in the range $0 < \omega < 1$.

   The Jacobi method with under relaxation is appropriate for application to solving the Newton equation.   
This method requires $O(n)$-time if the coefficient matrix is sparse (the number of non-zero elements 
in the matrix is $O(n)$) and the number of iterations is fixed.
The Jacobi method is an  algorithm of parallel type, and so   should be suitable for a hardware implementation
that can utilize its  massive parallelism. 

\subsection{Decoder architecture}

\begin{figure}[htbp]
\begin{center}
\includegraphics[scale=0.4]{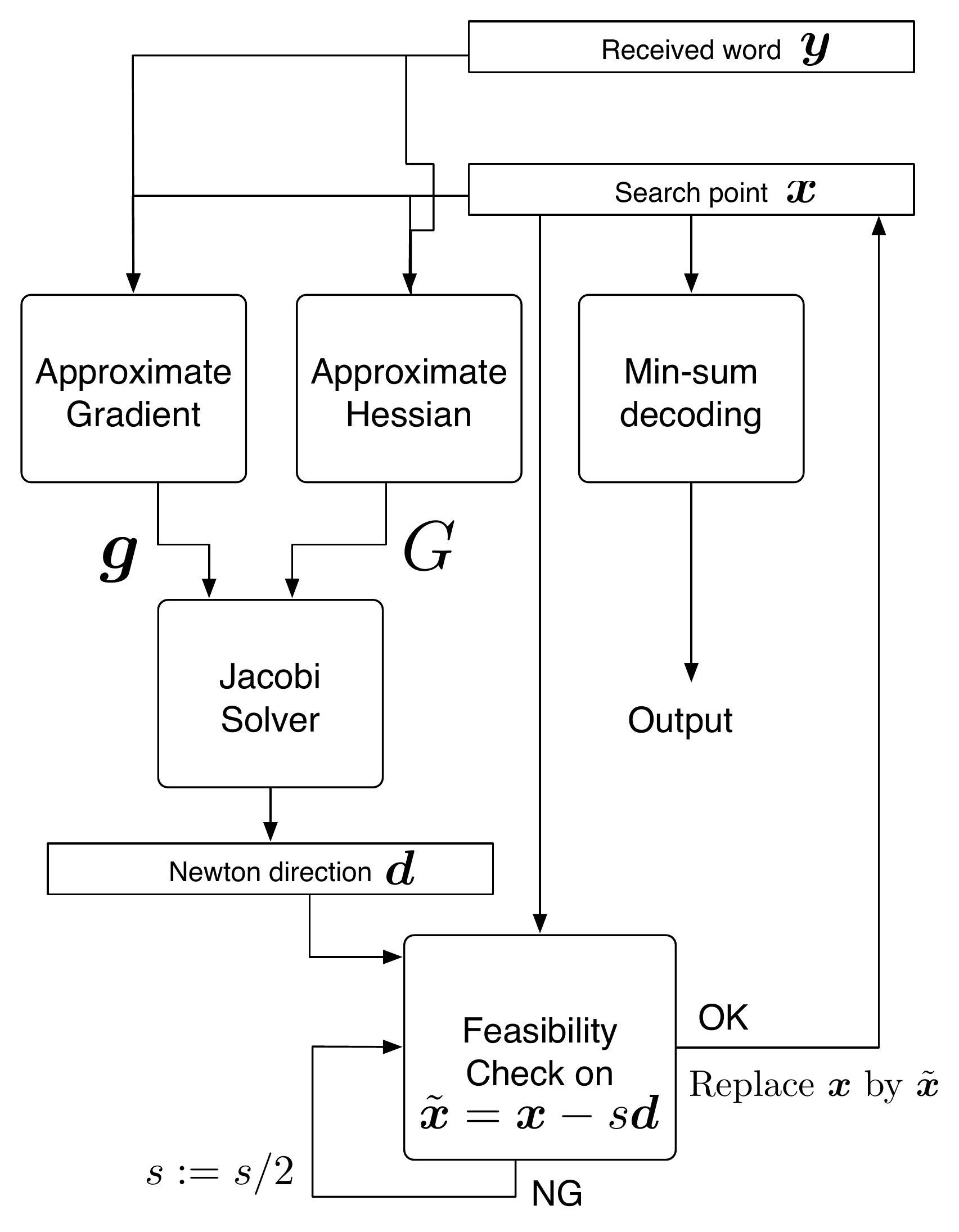} 
  \caption{A possible decoder architecture of the proposed interior point decoder.}
    \label{decoderarch}
\end{center}
\end{figure}

Figure \ref{decoderarch} shows a possible hardware architecture of the interior point decoder using the 
Newton method. There are five major processing blocks given by approximate gradient computation,
approximate Hessian computation, Jacobi solver, feasibility check, and built-in min-sum decoder.
Every block is suitable for parallel implementation. 
In order to design a high-speed decoder,   this parallelism must be exploited.

\section{Behavior of interior point decoding}

In this section, the behavior of the interior point decoding is discussed
on the basis of the results of computer simulations.
In the following simulations, a regular LDPC code with parameters  $n =204, m = 102, w_c = 3, w_r = 6$ 
is assumed where $w_c$ and $w_r$ denote  column and row weight, respectively.
The code is due to MacKay \cite{Encyclopedia}.
The channels used in the simulation are PR channels.

\subsection{Objective function values as a function of number of iterations}

%In this subsection, 
%we will see  the behaviors of the objective function values 
%as a function of the number of iterations.

\begin{figure}[htbp]
\begin{center}
\includegraphics[scale=0.7]{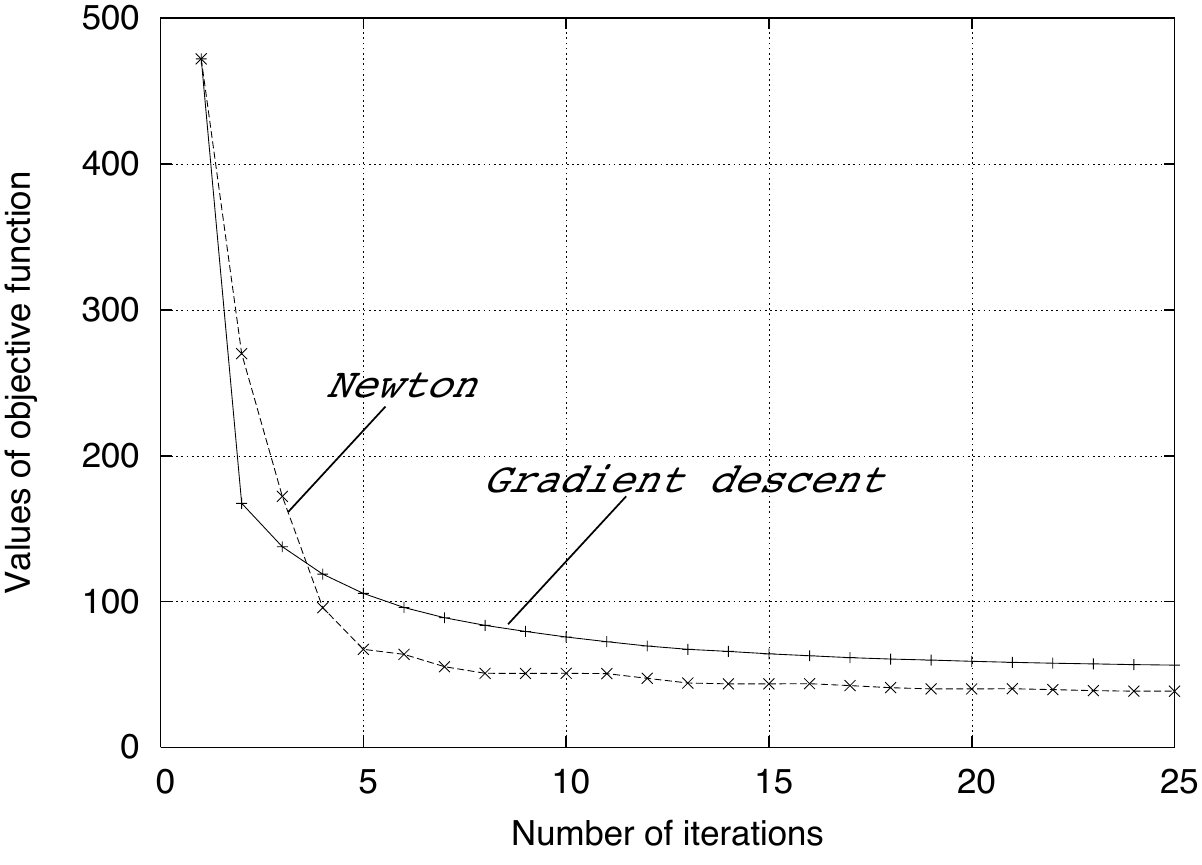} \\
$I_{max} = 5, O_{max} = 10, t_0 = 5.0, \alpha =2.0$,
$h_0 = 1.0, h_1 = -1.0, SNR = 8$ dB, average of 1000 trials  \\
  \caption{Values of the objective function: gradient descent and Newton methods.}
  \label{exp-meritcurve}
\end{center}
\end{figure}

Figure \ref{exp-meritcurve} presents the average values of the objective function $f(\ve x)$ as a
function of the number of iterations. The two curves in Fig.\ref{exp-meritcurve} correspond to 
the results of the proposed scheme obtained using the gradient descent-inner loop and
 the Newton-inner loop, respectively. 
The number of iterations is defined as the number of
executions of the inner loop in the decoding process.
These curves have been obtained by taking the average of 1000 trials (1000 codewords).

The parameters of the decoder are as follows:
$I_{max} = 5, O_{max} = 10, t_0 = 5.0, \alpha =2.0$.
The channel is the dicode channel (i.e., $h_0 = 1.0, h_1 = -1.0$) with an
SNR of $8$dB.
It may be observed that the average objective function values decreases rapidly 
in the first few iterations.
The curve of the  gradient descent method shows the fastest
decrement when the number of iterations is small (such as 1--3) .  
However, the Newton method (with Cholesky decomposition\footnote{Since no evident 
difference in
decoding performance 
between the Newton method with Cholesky decomposition and that with Jacobi method  has been observed, we assume the Newton method with Cholesky decomposition
throughout the section.}) gives 
smaller objective function values following the 4th iteration.
These results suggest that the proposed decoding algorithm using the Newton method 
may require fewer iterations compared with that employing the gradient descent method.

\begin{figure}[htbp]
\begin{center}
\includegraphics[scale=0.7]{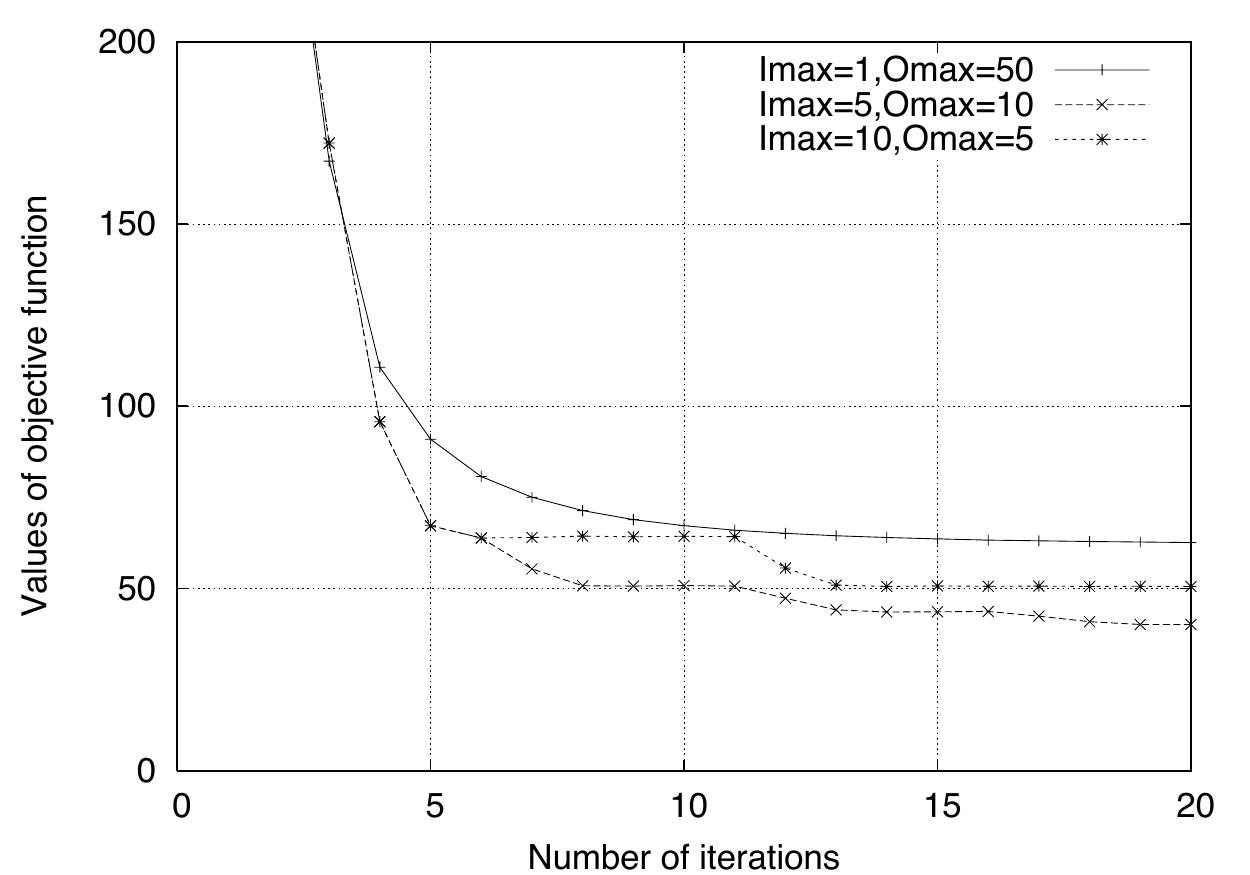} \\
$t_0 = 5.0, \alpha =2.0$, \\
$h_0 = 1.0, h_1 = -1.0, SNR = 8$ dB, average of 1000 trials  \\
  \caption{Balance of the number of inner and outer loops.}
  \label{exp-inner-outer}
\end{center}
\end{figure}

We next consider the balance of the number of inner/outer loops.
In the following experiment, the product of the number of inner and outer loops 
is set to be 50. Three combinations $(I_{max},O_{max}) = (1,50), (5,10), (10,5)$ have been tested, where
the interior point decoding using the Newton method with Cholesky decomposition has been
used.
Figure \ref{exp-inner-outer} presents the objective function curves for the three combinations, where
the parameters of the decoder are included in the figure.  
We can see that the pair $(I_{max},O_{max}) = (5,10)$ gives the smallest values after the
7th iteration.

\begin{figure}[htbp]
\begin{center}
\includegraphics[scale=0.7]{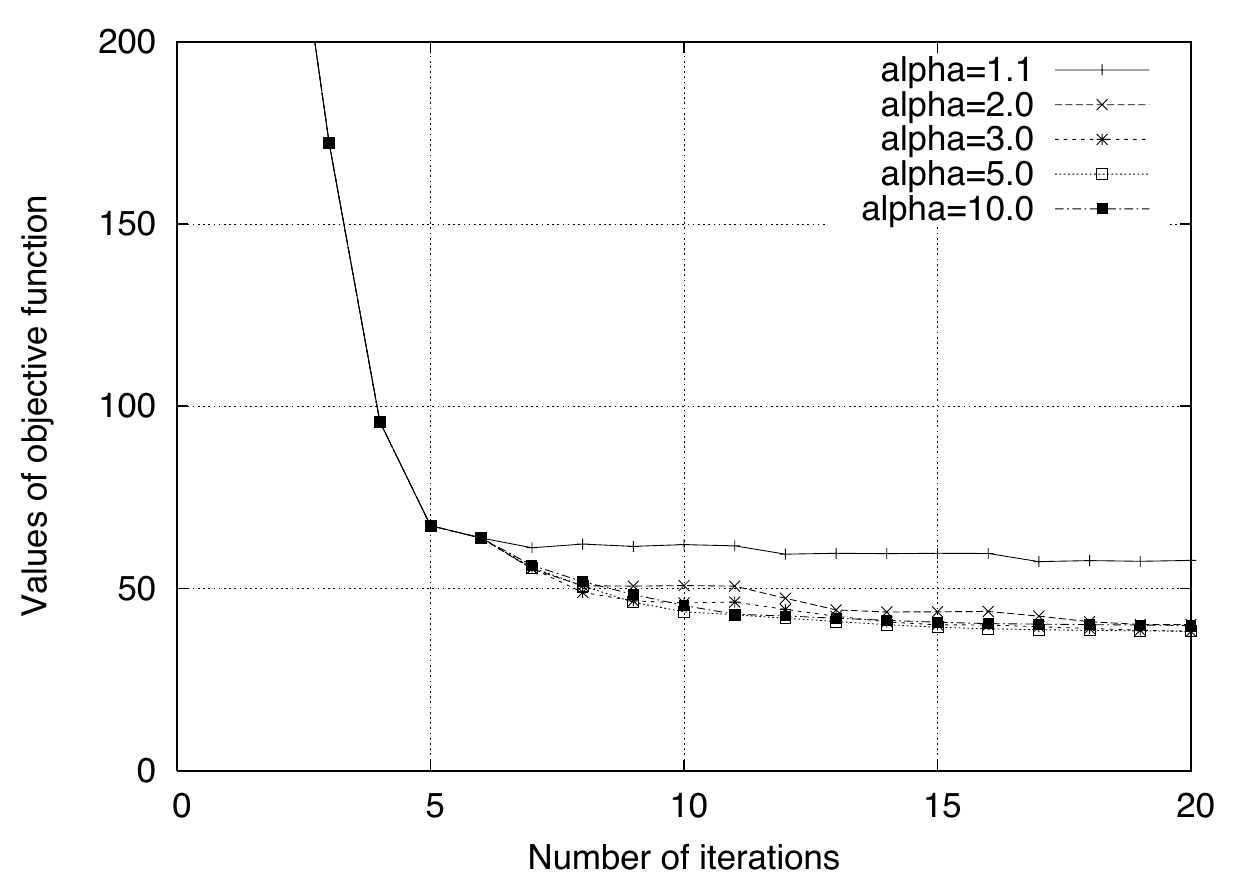} \\
$I_{max} = 5, O_{max} = 10, t_0 = 5.0$, \\
$h_0 = 1.0, h_1 = -1.0, SNR = 8$ dB, average of 1000 trials  \\
  \caption{Dependency of convergence on  scale factor $\alpha$.}
  \label{exp-alpha}
\end{center}
\end{figure}

Figure \ref{exp-alpha} plots the dependency on the scale parameter $\alpha$. 
Again, the interior point decoding using the Newton method with Cholesky decomposition is 
adopted. It is observed that the convergence properties of this method are not so sensitive to the choice of 
the scale factor $\alpha$. 

\subsection{Bit error probability of the interior point decoding}
In this subsection, the bit error probability curves of the interior point decoding 
obtained from computer simulations are presented. The code used in the simulations is
a regular LDPC code with parameters $n=204, m=102, w_c=3,w_r =6$.

\begin{figure}[htbp]
\begin{center}
\includegraphics[scale=0.45]{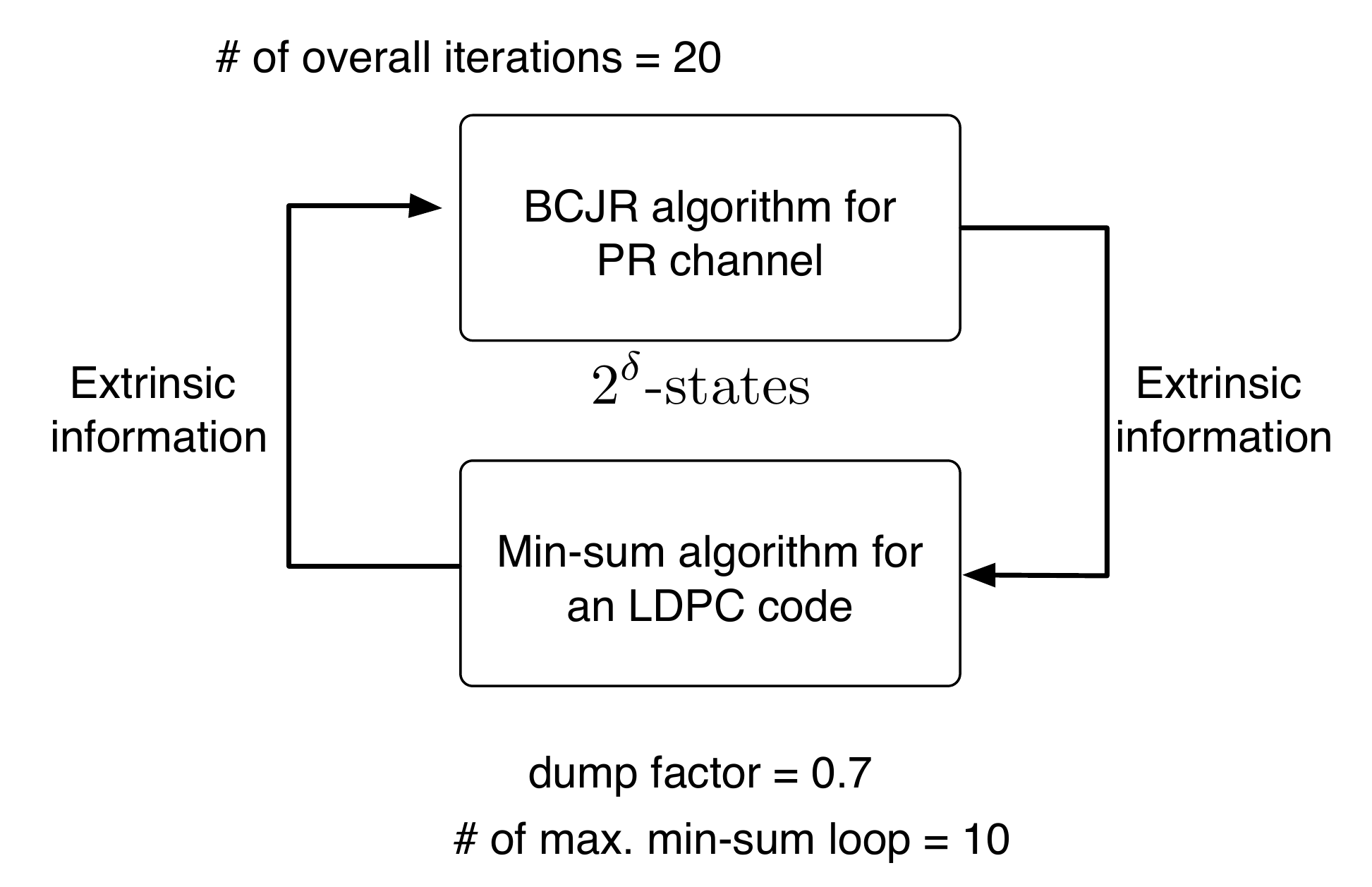} 
  \caption{Block diagram of joint message passing decoding (joint MPD).}
  \label{jointdecoding}
\end{center}
\end{figure}

In order to make a comparison with conventional decoding algorithms, we also obtained results using the joint 
message passing decoding (abbreviated as joint MPD) in this paper. The block diagram of 
the joint MPD is presented in Fig.\ref{jointdecoding}. The joint MPD consists of two parts: 
BCJR (Bahl, Cocke, Jelinek and Raviv) algorithm \cite{BCJR} for a PR channel 
and min-sum algorithm (with dump factor 0.7) for an LDPC code.
The BCJR algorithm computes extrinsic information in the standard way and passes this information to 
the min-sum algorithm. The min-sum algorithm uses the output from the BCJR algorithm as 
a priori information. The extrinsic information generated from the min-sum algorithm is then 
treated as a priori information in the BCJR algorithm. The two parts of this method are iteratively executed in turn.
We assume that the maximum number of iterations in the min-sum decoder is 10 and 
that the overall iterations are limited to 20. For each iteration, a parity check procedure is executed;
the iteration stops when the parity check passes.

\begin{figure}[htbp]
\begin{center}
\includegraphics[scale=0.7]{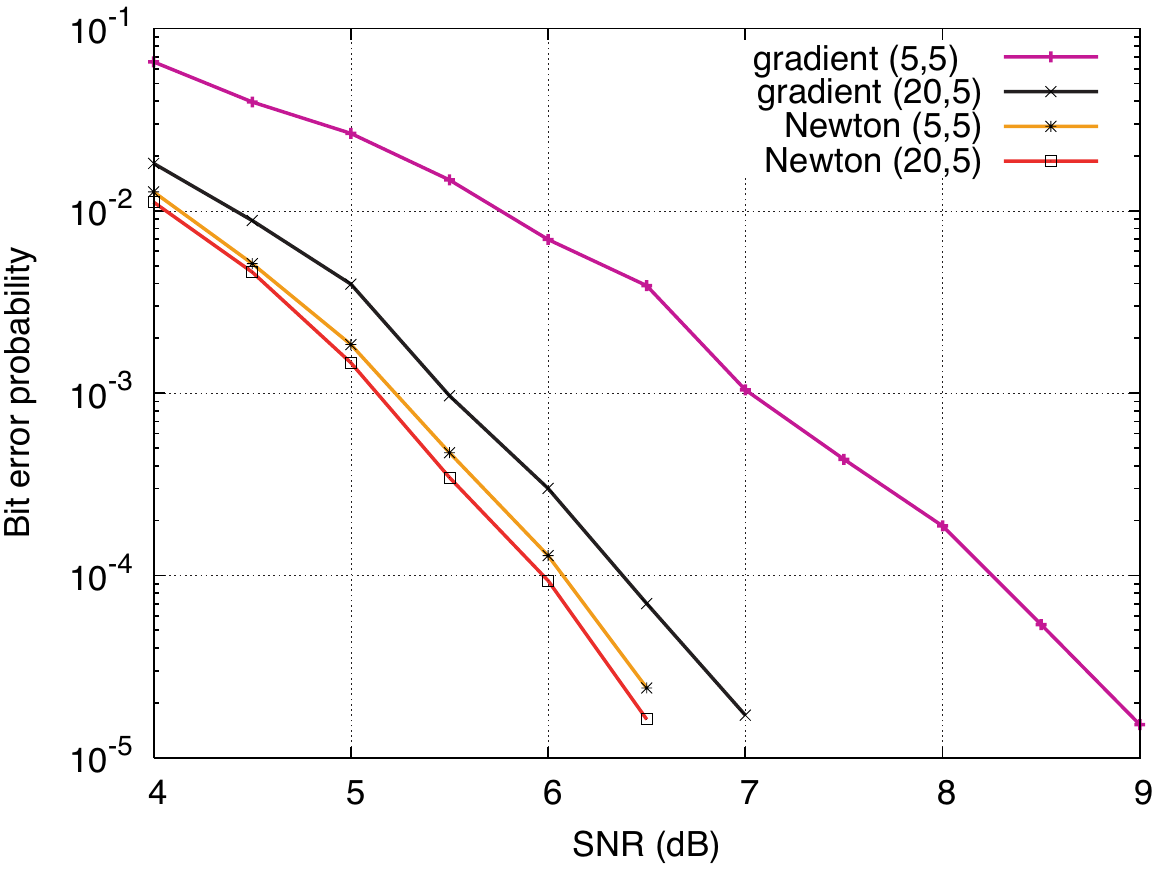} 
  \caption{BER curves of interior point decoding for  $1-D$ channel (1).}
  \label{exp-dicode1}
\end{center}
\end{figure}

Figure \ref{exp-dicode1} shows the BER (Bit Error Rate) curves for the
$1-D$ channel ($h_0 = h_1 = 1$). In this figure, the decoding performances of
the interior point decoding using the gradient descent and the Newton methods are compared.
For example, the label "gradient (5,5)" corresponds to the gradient descent method with 
$(I_{max}, O_{max})=(5,5)$. The parameters $t_0 = 5.0$ and $\alpha = 2.0$ are used in 
these simulations. The maximum number of iterations in the built-in min-sum decoder 
is $L_{max} = 20$ and the dump factor is $\kappa = 0.7$.
Comparing the gradient descent $(5,5)$ and 
the Newton $(5,5)$ results,   Newton (5,5) exhibits much smaller bit error probabilities
(approximately 2dB gain at BER = $10^{-3}$) than those of gradient (5,5). 
This difference could be a consequence of the faster convergence of the Newton method.

From Fig.\ref{exp-dicode1}, it can be observed that
a large improvement in BER can be obtained by increasing $I_{max}$ from 5 to 20 in the case of 
the gradient descent method.
On the other hand, only negligible improvement is achieved by increasing $I_{max}$ from 5 to 20
in the case of the Newton method.
From these observations and other simulation results, 
we may be able to conclude that the interior point decoding using the Newton method requires at least 
5 inner-iterations to achieve most of the potential performance. 
By contrast, the interior point decoding using the gradient descent method requires at least 20 iterations
(however, more   than 20 iterations offers only a marginal improvement).

\begin{figure}[htbp]
\begin{center}
\includegraphics[scale=0.7]{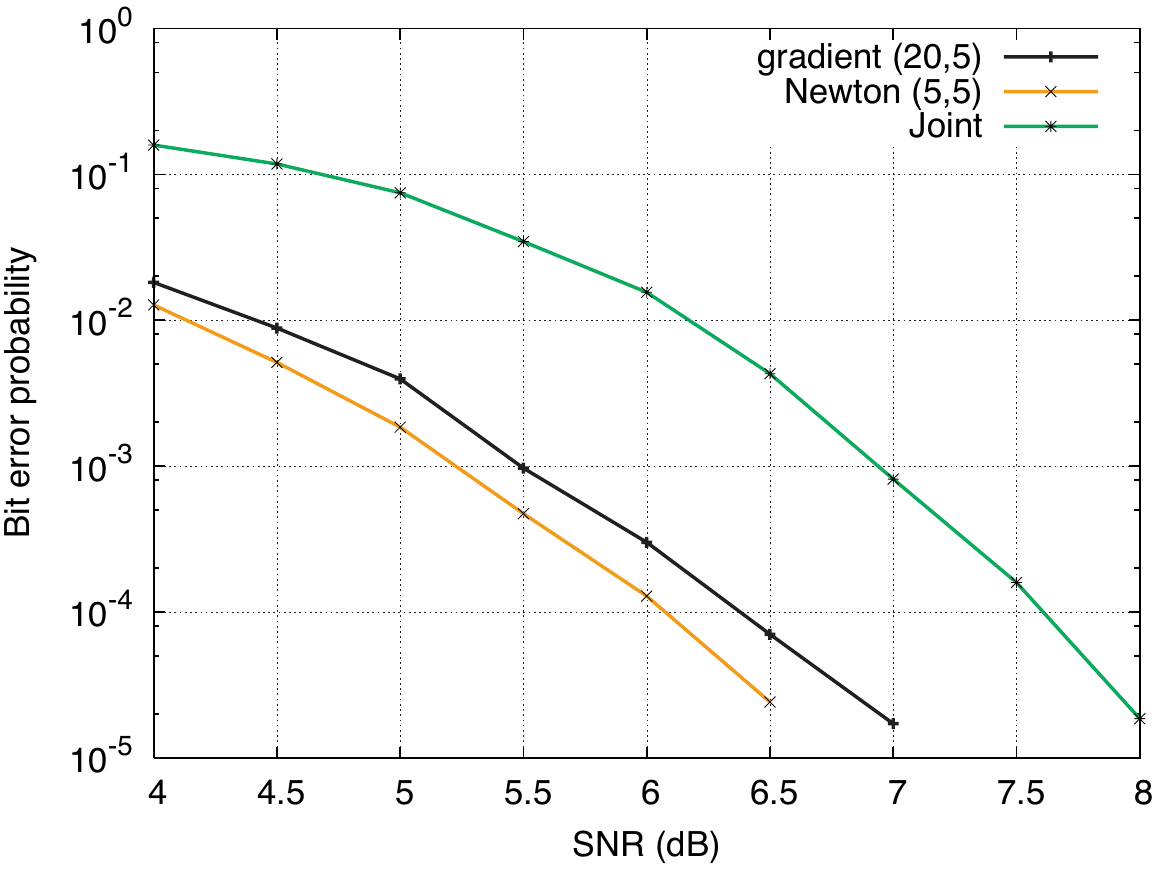} 
  \caption{BER curves of interior point decoding for $1-D$ channel (2).}
  \label{exp-dicode2}
\end{center}
\end{figure}

Figure \ref{exp-dicode2} presents BER curves of gradient (20,5), Newton (5,5), and joint MPD.
The channel is also a $1-D$ channel, as in the case of Fig.\ref{exp-dicode1}.
From Fig.\ref{exp-dicode2}, firstly, we can see that interior point decoding yields a better decoding 
performance than that of joint MPD. The difference is approximately 1.5dB (Newton v.s. joint MPD) 
at bit error probability $10^{-4}$. Secondly, it is observed that Newton method gives smaller 
bit error probabilities than those obtained by using the gradient descent method.

\begin{figure}[htbp]
\begin{center}
\includegraphics[scale=0.7]{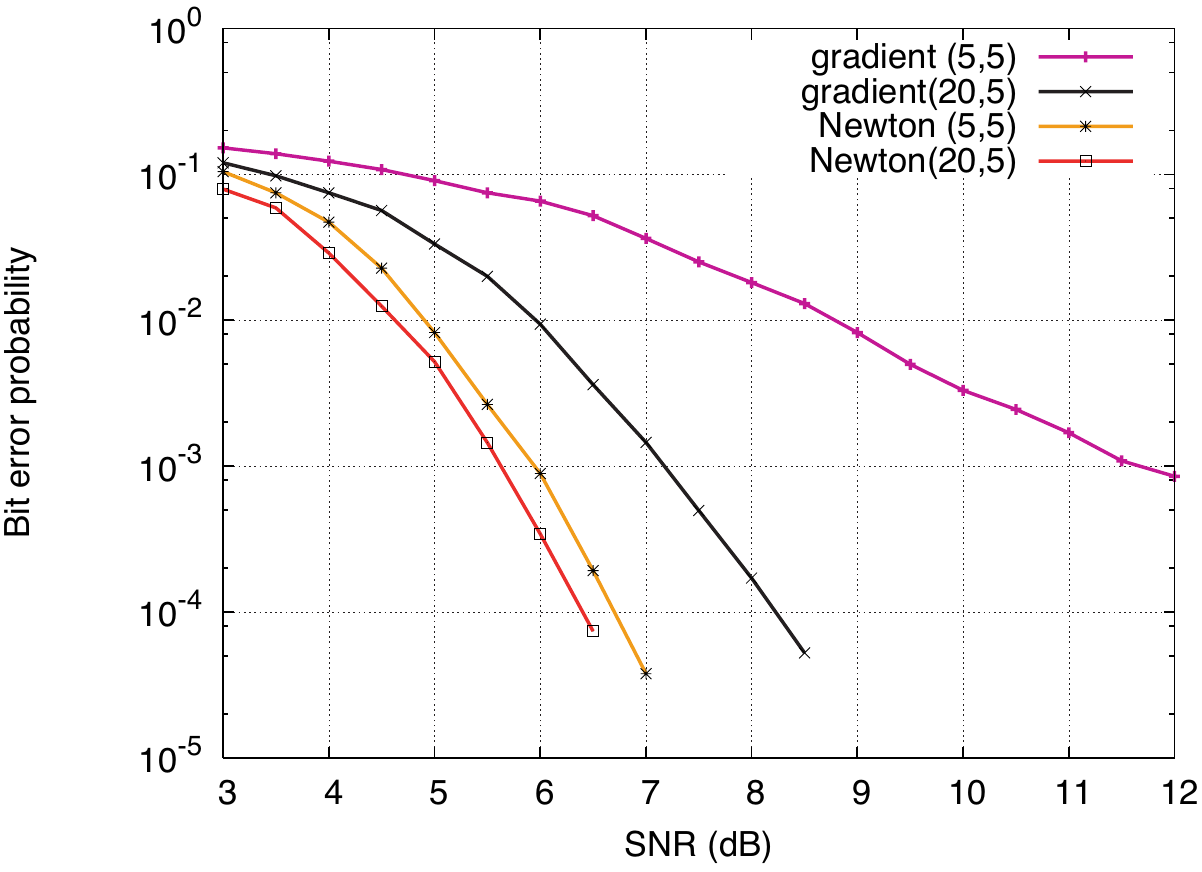} 
  \caption{BER curves of interior point decoding  for a long tail PR channel $(\delta = 16)$.}
  \label{exp-longtail}
\end{center}
\end{figure}

Figure \ref{exp-longtail} shows BER curves for a long tail PR channel $(\delta = 16)$.
Note that joint MPD cannot be applied to such a PR channel with large number of 
states (i.e., PR channel with long tail PR coefficients) because of 
the huge computational costs of BCJR computation, and so  Fig.\ref{exp-longtail} does not
include the BER curve for joint MPD.
The PR coefficients of this long tail PR channel are
\begin{eqnarray} \nonumber
&&\hspace{-5mm}\{ h_0,h_2,\ldots,h_{16} \} = \\ \nonumber
&&\hspace{-5mm}\begin{array}{rrrrrr}
\{1.000, & 0.253, &-0.293,&0.084, &-0.057,& 0.992,   \\
 -1.438, &-0.910,&0.106, &-0.600,&-0.844, &0.018,   \\ 
  0.197, &-0.743,&0.490, &-0.070,&1.43 \}.   &           \\
\end{array}
\end{eqnarray}
The coefficients $\{h_1,\ldots h_{16} \}$ are samples of a Gaussian random variable 
with mean 0 and variance 1. We can see that the interior point decoding certainly has the
capability to decode a given received vector observed from such a long tail PR channel.  This 
can be considered as an advantage of the interior point decoding.
Comparing the results of the gradient (20,5) and Newton (5,5) methods, the latter gains approximately 1.5 dB gain  at a bit error probability of $10^{-5}$.

\begin{figure}[htbp]
\begin{center}
\includegraphics[scale=0.7]{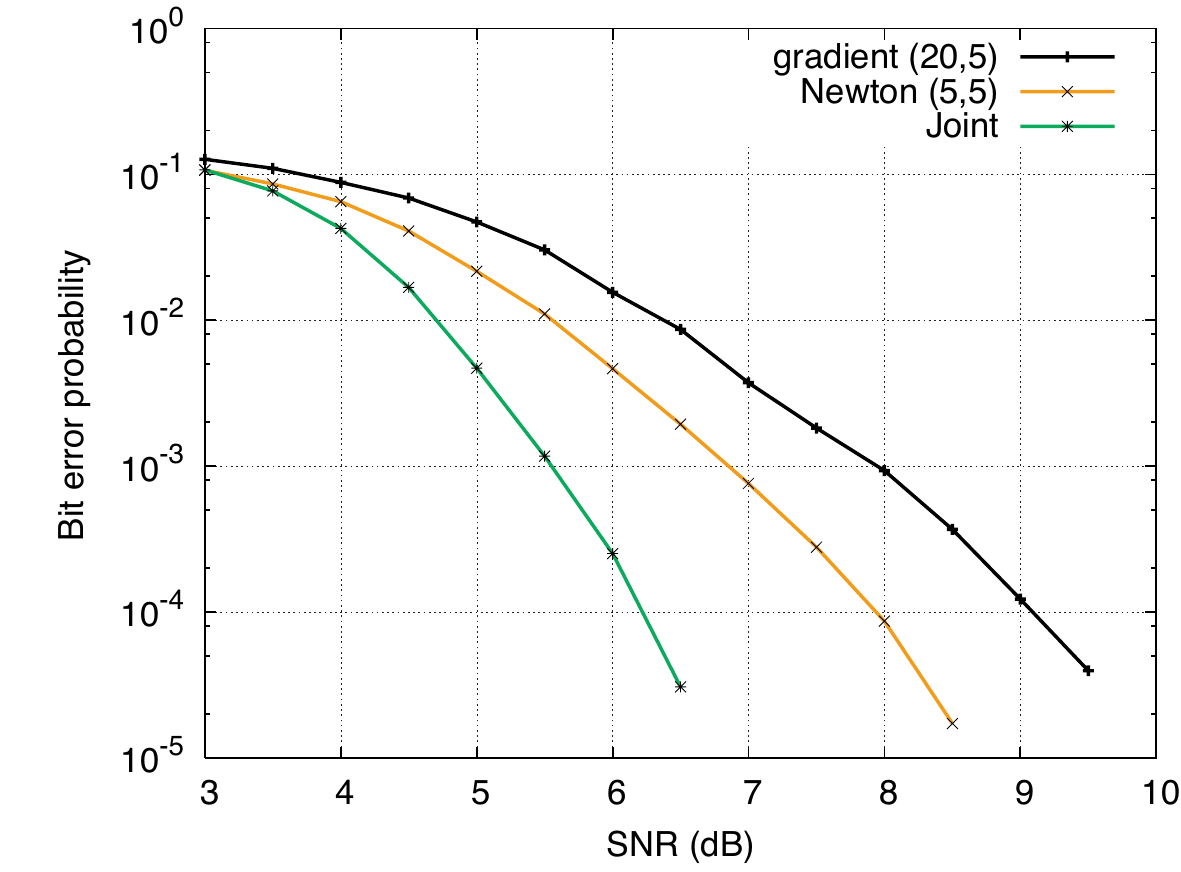} 
  \caption{BER curves of interior point decoding for an EPR4 channel $(h_0=h_1=1, h_2 = h_3=-1)$.}
  \label{exp-epr4}
\end{center}
\end{figure}

Figure \ref{exp-epr4} deals with the case of an EPR4 channel $(\delta = 3, h_0=h_1=1, h_2 = h_3=-1)$.
In this case, among three decoding algorithms (gradient descent (20,5), Newton (5,5), and joint MPD), 
joint MPD yields the best decoding performance. A large gap (1.8dB at BER $=10^{-4}$) exists between the
BER curves of joint MPD and Newton (5,5).  This performance degradation may be explained from the
geometrical view point. Some vertices of the mapped polytope associated with this channel would have a
very thin decision region. 

\begin{figure}[htbp]
\begin{center}
\includegraphics[scale=0.7]{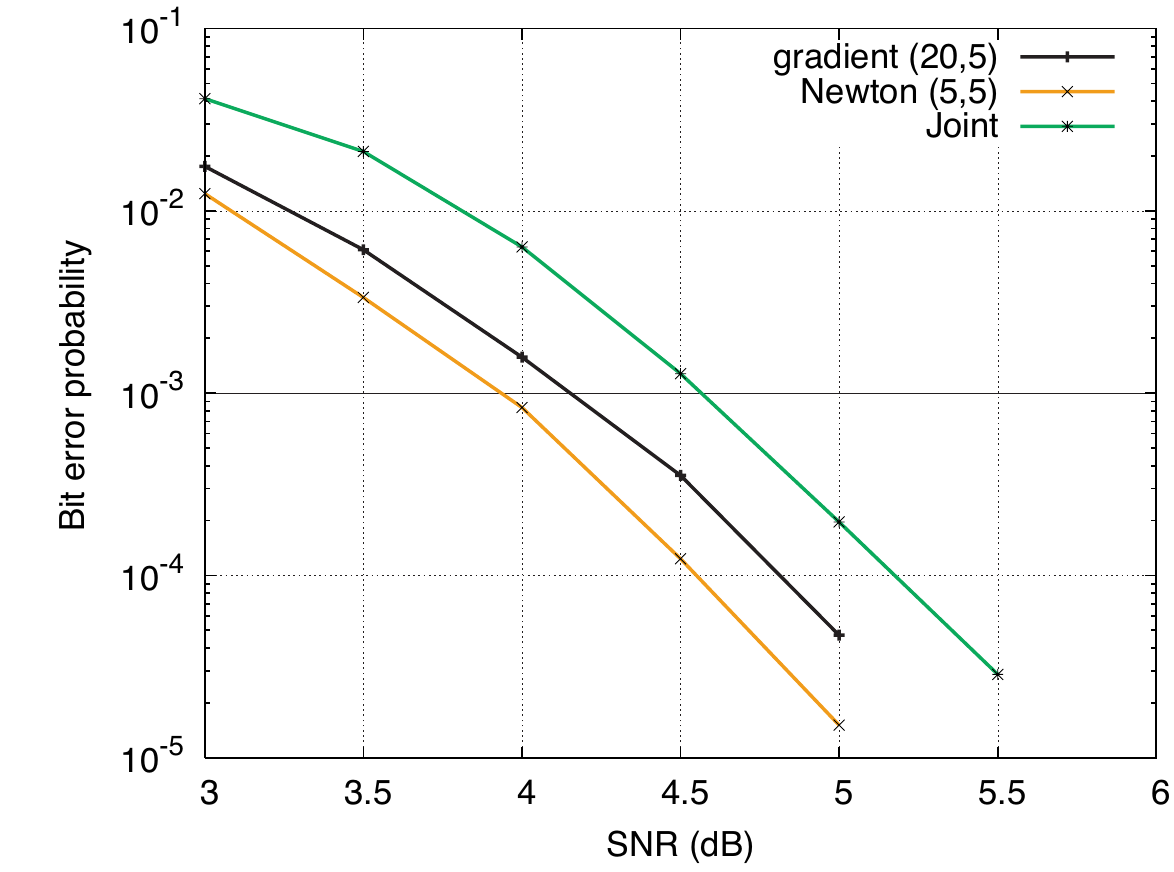} 
  \caption{BER curves of interior point decoding for another $\delta=3$ channel $(\delta = 3, h_0=h_1=h_2 =1, h_3=-1)$.}
  \label{exp-deg3}
\end{center}
\end{figure}

Figure \ref{exp-deg3} presents the case where the PR coefficients are  $(h_0=h_1=h_2 =1, h_3=-1)$. This 
channel and the EPR4 channel have the same degree, $\delta = 3$, with the channels differing  only in the sign of the coefficient $h_2$.
From Fig.\ref{exp-deg3}, we can see that interior point decoding offers smaller bit error probabilities than those of 
joint MPD across the entire range of SNR.  These results suggest that the decoding performance of the interior point decoding
is highly dependent on the PR coefficients of the channel.

Table \ref{throughput} presents the throughput of the software implemented decoders.
Here the throughput of the decoder is defined as the number of codewords tested in 
a second. Namely, it includes the time for encoding, noise generation, and decoding.
It is fair to say that the values of throughput highly depend on the implementation and
thus it should be considered as rough estimates of the decoding complexity.
From Table \ref{throughput}, we can observe that the proposed algorithm 
gives a higher throughput than that of joint MPD. 
It is also interesting to see that 
Newton (5,5) achieves a much higher throughput than that of gradient descent (20,5).
This improvement on throughput  is mainly due to  the faster convergence  of the Newton method, which can compensate the additional complexity (solving the Newton equation) required by this method.
Note that a throughput of 1715 blocks/sec corresponds to 583 $\mu$ second per codeword.
\begin{table}
\begin{center}
\caption{Throughput of Software Implemented Decoders.}
\label{throughput}
\begin{tabular}{ll}
\hline 
\hline 
decoder & throughput (blocks/sec) \\
\hline 
gradient descent (20,5) &  706 \\
Newton with Cholesky (5,5) & 1715 \\
Newton with Jacobi (5,5) & 1506 \\
\hline
joint MPD & 359 \\
\hline
\end{tabular} \\
Channel: $1-D$ channel, SNR = 6 dB \\
Code:  $n=204, m=102, w_c = 3, w_r = 6$ \\
Computer environment: Mac Pro with intel Xeon 2.0 GHz
\end{center}
\end{table}

\section{Conclusion}

In this paper, the interior point decoding for linear vector channels is presented.
The proposed algorithm is based on the principle of a relaxed MLD rule which 
is a convex optimization problem.
Approximate variations of the gradient descent and the Newton methods 
play a key role in the interior point algorithm  to solve the convex optimization problem. 
Several efficient implementation 
techniques have been developed in order to realize a decoder with reasonable 
computational complexity.
The proposed algorithm may be applicable to various kinds of channels with memory,
such as channels with additive correlated noise, MIMO channels, and 2D-ISI channels. 

Error analysis based on the geometrical properties of the fundamental polytope is given as well.
The decision regions of a relaxed ML decoder can be characterized by the normal cones
of the affine image of the fundamental polytope. 
This geometrical view  helps us to understand the behavior of a sub-optimal relaxed ML decoder based on
convex optimization.

As a matter of fact, we cannot  simply conclude that the optimization approach is superior to the message passing
approach. There are some cases where joint MPD overcomes the 
proposed algorithm (e.g, EPR4 case).  Furthermore, there exist other configurations of
a joint MPD (see \cite{Brian}) which have not been tested in this paper.

However, the simulation results presented in the previous section 
are encouraging and they show the potential of the optimization approach.
Compared with a conventional joint MPD, 
the proposed decoding algorithm achieves better BER 
performance with less decoding complexity in the case of PR channels in many cases.
An advantage of the proposed algorithm is that it is capable of handling the channels with long memory.

The present paper discusses a scheme based on a relatively simple interior point 
algorithm (i.e., barrier function method).
More sophisticated convex optimization algorithms with faster convergence property \cite{Boyd} \cite{Bert}
can be considered in future.
For example, recently, Vontobel \cite{Vontobel} presented 
a new interior point algorithm for linear programming decoding.
The algorithm is based on the primal-dual interior point algorithm which appears promising in terms of
the convergence speed.

The efficient implementation techniques (fast feasibility check, approximate 
gradient, approximate Hessian) developed in this paper and the formulation of the barrier function 
representing the fundamental polytope  
could be useful not only in the proposed algorithm but also in forthcoming 
decoding algorithms based on various types of the interior point algorithm.

\section*{Acknowledgment}
 This work was supported by the Ministry of Education, Science, Sports
and Culture, Japan, Grant-in-Aid for Scientific Research on Priority Areas
(Deepening and Expansion of Statistical Informatics) 180790091
and a research grant from SRC (Storage Research Consortium). 
%The author wish to express his sincere thanks to anonymous reviewers whose
%comments help to improve the quality of the paper.

\end{document}